\documentclass[manuscript]{aastex}
\usepackage{epstopdf}
\usepackage{subfigure}

\usepackage{longtable}

\shorttitle{Nanoflare trains}
\shortauthors{Reep, Bradshaw, \& Klimchuk}

\begin{document}

\title{Diagnosing the time-dependence of active region core heating from the emission measure:  II. Nanoflare trains}

\author{J. W. Reep}
\affil{Department of Physics and Astronomy, Rice University, Houston, TX 77005, USA}
\email{jeffrey.reep@rice.edu}\and
\author{S. J. Bradshaw}
\affil{Department of Physics and Astronomy, Rice University, Houston, TX 77005, USA}
\email{stephen.bradshaw@rice.edu}\and
\author{J. A. Klimchuk}
\affil{NASA Goddard Space Flight Center, Solar Physics Lab., Code 671, 8800 Greenbelt Road, Greenbelt, MD 20771, USA}
\email{james.a.klimchuk@nasa.gov}

\begin{abstract}
The time-dependence of heating in solar active regions can be studied by analyzing the slope of the emission measure distribution cool-ward of the peak. In a previous study we showed that low-frequency heating can account for 0\% to 77\% of active region core emission measures. We now turn our attention to heating by a finite succession of impulsive events for which the timescale between events on a single magnetic strand is shorter than the cooling timescale. We refer to this scenario as a ``nanoflare train'' and explore a parameter space of heating and coronal loop properties with a hydrodynamic model. Our conclusions are: (1) nanoflare trains are consistent with 86\% to 100\% of observed active region cores when uncertainties in the atomic data are properly accounted for; (2) steeper slopes are found for larger values of the ratio of the train duration $\Delta_H$ to the post-train cooling and draining timescale $\Delta_C$, where $\Delta_H$ depends on the number of heating events, the event duration and the time interval between successive events ($\tau_C$); (3) $\tau_C$ may be diagnosed from the width of the hot component of the emission measure provided that the temperature bins are much smaller than 0.1~dex; (4) the slope of the emission measure alone is not sufficient to provide information about any timescale associated with heating - the length and density of the heated structure must be measured for $\Delta_H$ to be uniquely extracted from the ratio $\Delta_H/\Delta_C$.
\end{abstract}


\section{Introduction}
\label{introduction}

It remains a difficult question whether the solar corona, at temperatures exceeding a million degrees, is heated steadily or impulsively.  Any potential heating mechanism must explain observed extreme ultraviolet (EUV) and X-ray emission, as well as observed dynamical activity.  We must understand the mechanism, how it stores and releases energy, how that energy release affects the plasma and produces the emission, and finally we must predict observable properties of coronal loops \citep[see][]{klimchuk2006}.  Although we cannot directly observe coronal heating, we can study the time dependence of the heating which may give insight into the mechanism by which energy is released.

Impulsive heating is characterized by short bursts of energy release \citep[which we refer to as nanoflares; see, for example,][]{cargill1997} and a period of cooling between successive heating events.  If the cooling period is short then the plasma will not cool much between heating events and the heating may essentially be treated as steady (the limit of high-frequency nanoflares), resulting in a more or less isothermal plasma distribution. Contrariwise, if successive bursts are sufficiently far apart then the plasma will cool significantly and, at any particular time, plasma on different magnetic strands of a multi-stranded loop may have a broad distribution of temperatures.  Thus, the temperature distribution of plasma, quantified by the emission measure (EM), may help to distinguish between different heating scenarios. We will focus on heating in active region cores in the present work.

As shown in the first paper of this series \cite{bradshaw2012} (hereafter referred to as Paper~I), EMs in active region (AR) cores can be characterized by the slope $\alpha$ (i.e. $EM \propto T^{\alpha}$) between the temperature of peak emission (typically around 3-5 MK) and 1~MK, with observed values ranging from 1.70 to 5.17 \citep{warren2011, winebarger2011, tripathi2011, warren2012, schmelz2012}.  This slope measures the amount of warm plasma ($T \approx 1$~MK) relative to the amount of hot plasma ($T>3$~MK).  Further, the distribution falls off with a very steep slope at temperatures above the EM peak; that is, there is very little observable emission hotter than at the peak because the emission measure of the plasma is low and because of strong non-equilibrium ionization effects \citep{bradshaw2006,reale2008}.

In Paper~I we found that the large uncertainties in observed $\alpha$ values yield as many as 77\% or as few as zero observed active region EMs that are consistent with low-frequency nanoflare heating and that low-frequency nanoflares cannot explain the upper range of observed EM slopes. \cite{warren2011} found that very high-frequency heating leads to hot, isothermal EMs, which have not been observed. Warm emission is also present in AR cores \citep[e.g.][]{viall2011,viall2012,warren2012}. The key to obtaining steeper slopes is to enhance the amount of hot emission relative to the amount of warm emission, but one must still account for the presence of warm emission. Furthermore, the hot component of the EM is itself not generally perfectly isothermal and can extend into temperature bins either side of the peak. Increasing the frequency of heating events will yield steeper EM slopes, but it cannot be increased so much that the resulting EM is effectively isothermal, and the duration of heating cannot be so long that no warm emission is ever produced.

These requirements have led us to explore the possibility of what we term ``nanoflare trains'', which we define as nanoflares that occur on a single magnetic strand and repeat at intermediate frequencies. An intermediate frequency is such that the time interval between successive heating events is less than the cooling timescale so that another nanoflare occurs on the same strand before the loop cools fully, but is not so high that the heating is effectively steady. The heating eventually ceases so that the loop then cools and drains. Intermediate-frequency heating can maintain the plasma at high temperature for longer than low-frequency heating, which enhances the amount of hot emission relative to the amount of warm emission produced when the plasma finally cools, and steepens the EM slope. These nanoflare train properties can therefore satisfy the requirements of yielding steeper slopes, ensuring that the hot component of the EM has some intrinsic width by allowing a period of cooling between successive heating events \citep[e.g.][found that EM peaks in ARs indicate conditions in which the inter-event cadence is shorter or of order the plasma cooling time]{susino2010}, and accounting for the presence of warm plasma by allowing the strands to cool fully following the cessation of the nanoflare train. It is well known that warm plasma is overdense compared to hydrostatic equilibrium and cooling provides a natural explanation for this observation \citep{warren2002,spadaro2003}.

We note here that steep EM slopes can also be obtained in a scenario of constant heating followed by full cooling and draining when the heating is switched off. The longer the heating is applied, the more hot emission relative to warm emission and the steeper the EM slope. This scenario is equivalent to a high-frequency nanoflare train \citep[e.g.][]{warren2010}. While this scenario is feasible, it is not one that we favor for three reasons: (1) it is difficult to imagine a scenario in which the power delivered to the plasma remains constant and smooth, even if it were continuous we may expect it to fluctuate which could look like individual pulses - the very process of releasing energy alters the properties of the magnetic field and the plasma; (2) the hot component of the EM would appear much more isothermal than it does; and (3) continual, small-scale bursts of activity observed at the limits of current instrument resolution support frequent, short timescale events. All feasible theories of coronal heating to-date predict that individual magnetic strands are heated in an impulsive manner. This includes both magnetic reconnection-like processes and wave heating \citep{klimchuk2006}.

We will present the results of sixty numerical simulations of nanoflare train heating, exploring a wide parameter space, in an effort to explain the properties of observed EM distributions and we will show that an observational bias in measured $\alpha$ values may arise depending upon the EM reconstruction method employed. In Section~\ref{modeling} we describe the numerical aspects of this work and in Section~\ref{results} we present and discuss our findings. We summarize our results, present our key conclusions and discuss directions for future work in Section~\ref{summary}.

\section{Numerical modeling}
\label{modeling}

The current work presents the results of numerical calculations performed to investigate the feasibility of coronal loop heating by nanoflare trains. The calculations were carried out using the HYDRAD code (Bradshaw \& Mason 2003; Bradshaw \& Klimchuk 2011; Paper~I), which solves the one-dimensional equations of hydrodynamics that describe the behavior of a two-fluid plasma confined to an isolated magnetic strand.  The primary equations (conservation of mass, momentum, and energy) and associated assumptions are described in \cite{bradshaw2011}. The particular details of the calculations carried out here are summarized in Paper~I, with the exception of the implementation of heating by nanoflare trains which we will discuss now.

As in Paper~I we assume that the heating arises from the impulsive release of energy from the magnetic field and so the total amount of energy available to heat the plasma must be limited to the amount of free energy in the field. We preferentially energize the electrons and make the assumption that not all of the free energy is released from the field during a single heating event, but instead during a series of nanoflares: a nanoflare train. The free magnetic energy density is given by

\begin{equation}
E_B = \frac{\left(\epsilon B_p\right)^2}{8\pi},
\label{eqn1}
\end{equation}

\noindent where $B_p$ is the potential component of the field and $\epsilon$ parameterizes the level of stress such that $B_s = \epsilon B_p$ is the stress component. We choose $\epsilon=0.3$ (Dahlburg et al. 2005; Paper~I). \cite{mandrini2000} studied how the average values of $B$ and $B^2$ depend on the length of the field line, $2L$. Table~\ref{table1} gives results for several observed active regions based on Equation~9 of their paper.

\begin{table}
\centering
\caption{Average magnetic field strength $\left(<\!B^2\!>^{1/2}\right)$, free energy and volumetric heating rate versus loop length.}
\begin{tabular}{c c c c c c}
\tableline
$2L$ & $B_{\mbox{min}}$  & $B_{\mbox{max}}$ & $B_{\mbox{avg}}$ & $E_B$ & $E_{H0}$ \\
 $[$Mm] & [G] & [G] & [G] & [erg cm$^{-3}$] & [erg cm$^{-3}$ s$^{-1}$] \\ \tableline
40 & 83  & 189 & 136 & 66.23 & 0.03680 \\
80 & 42  & 150 &  94 & 31.64 & 0.01760 \\
160 & 18 & 89 &  51 & 9.31 & 0.00517 \\
\end{tabular}
\label{table1}
\end{table}

\begin{figure}
\centering
\includegraphics{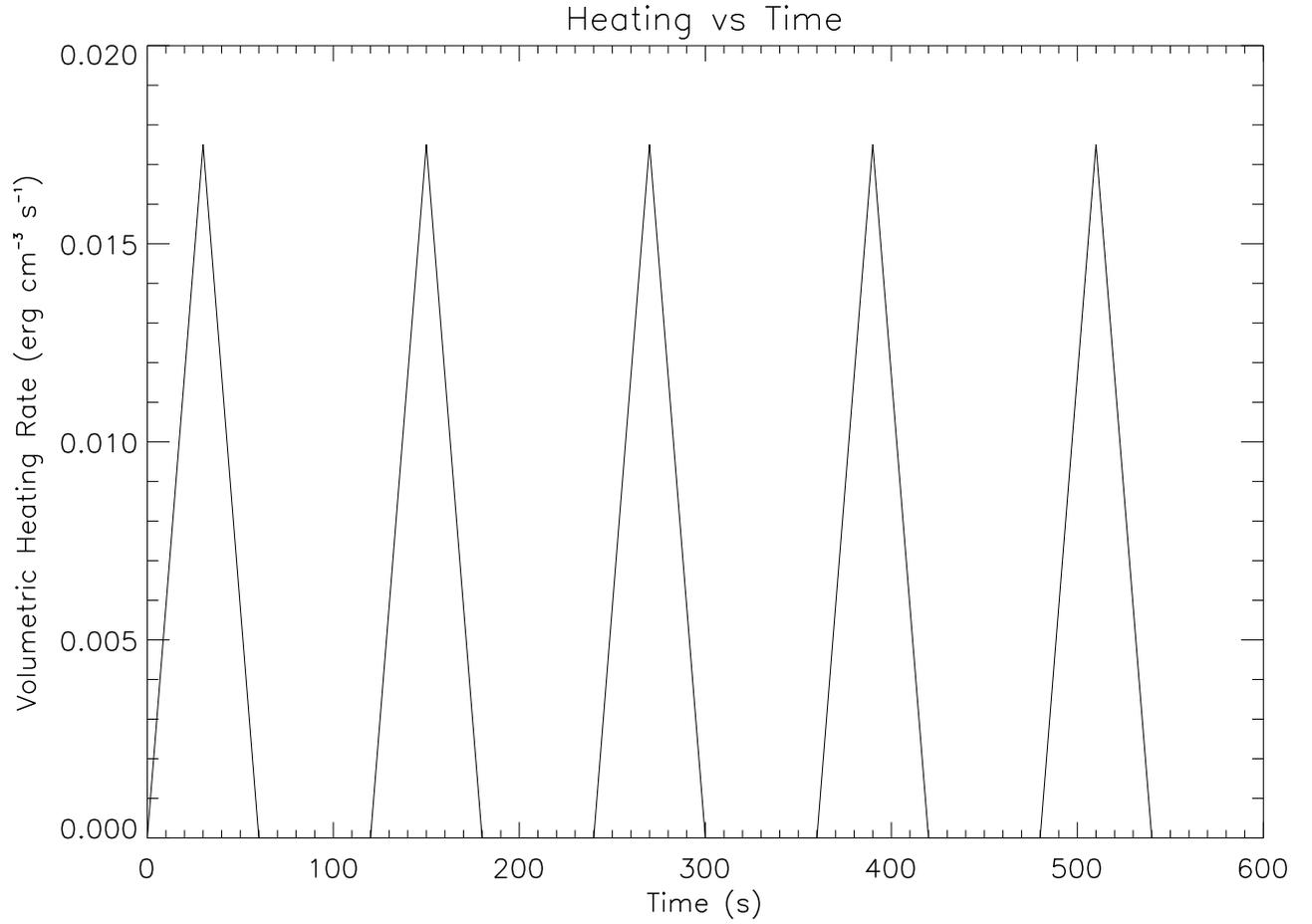}
\caption{The temporal heating profile of the nanoflare train for Run~21, where $2L=80$~Mm, $N=5$, $\tau_H=60$~s and $\tau_C=60$~s.}
\label{fig1}
\end{figure}

Since $B_{\mbox{min}}$ and $B_{\mbox{max}}$ can vary quite substantially we have assumed that $B_p \approx B_{\mbox{avg}}$ in our calculation of the free magnetic energy density in the fifth column of Table~\ref{table1}. This is the total energy available for release during the nanoflare train. To determine the peak volumetric heating rate of energy release $E_{H0}$ for each nanoflare of the train we need to know the length of the field line $2L$ (so that we can find $B_{\mbox{avg}}$ and hence $E_B$), the number of nanoflares $N$ comprising each train, and the duration of each individual heating event $\tau_H$.

We have chosen to explore a parameter space in which $2L=[40,80,160]$~Mm, $N=[5,10,15,20]$ and $\tau_H=[60,180,300]$~s, and carry out 3 sets of 20 numerical experiments; 1 set for each value of $2L$. The range of $\tau_H$ was decided by considering what constitutes an impulsive heating event. The most straightforward way to define impulsive heating is to set an upper limit to the timescale such that $\tau_H << \Delta_C$, where $\Delta_C$ is the total cooling timescale (thermal, radiative, enthalpy-driven) at the cessation of heating. Simulations of the secondary instability of electric current sheets indicate heating timescales of order 100~s, though this is highly variable depending on the thickness of the sheet \citep{dahlburg2005}. For the range of loop lengths, temperatures and densities typically encountered in active regions, $\Delta_C$ is in the region of a few thousand seconds (Table~\ref{table3}). We have set what we consider to be a reasonable limit to impulsive heating of about 10\% of this timescale and investigate a range of $\tau_H$ within this limit. However, we note here that longer duration heating may be appropriate for the very longest loops (160~Mm) in our study and the consequences of increasing $\tau_H$ for longer loops can easily be inferred from our results for shorter loops. We chose the range values of $N$ in order to create nanoflare trains with total durations $\Delta_H$ that are both shorter and longer than $\Delta_C$, so that we could investigate in detail the relationship between the amount of hot plasma (dependent on $\Delta_H$), the amount of cooling plasma (dependent on $\Delta_C$) and the EM slope. Another factor in the choice of $N$ is that the loop lifetime should be consistent with observed lifetimes on the order of hours. We have also chosen to keep $E_{H0}$ constant for each set of 20 experiments, so that we can focus on the influence of $N$, $\tau_H$ and $\tau_C$ (the time interval between successive events) on $\alpha$ $\left( \mbox{where }EM(T) \propto T^\alpha \right)$. Assuming that the free energy is divided more or less equally among each of the individual nanoflares and the temporal profile of the heating is triangular (a linear rise and decay, e.g. Figure~\ref{fig1}), the total energy release per unit volume can be found from

\begin{equation}
E_H = \frac{1}{2} N \tau_H E_{H0}.
\label{eqn2}
\end{equation}

\noindent Setting $E_H = E_B$ and choosing $N=20$ and $\tau_H=180$~s, we find the values of $E_{H0}$ listed in the final column of Table~\ref{table1}. This choice of $E_{H0}$ ensures that for $\tau_H=300$~s and all $N$, $B_{\mbox{min}} < B < B_{\mbox{max}}$ (where $B$ is the field strength implied by $E_H$).

\begin{figure}
\centering
\includegraphics[width=0.8\textwidth]{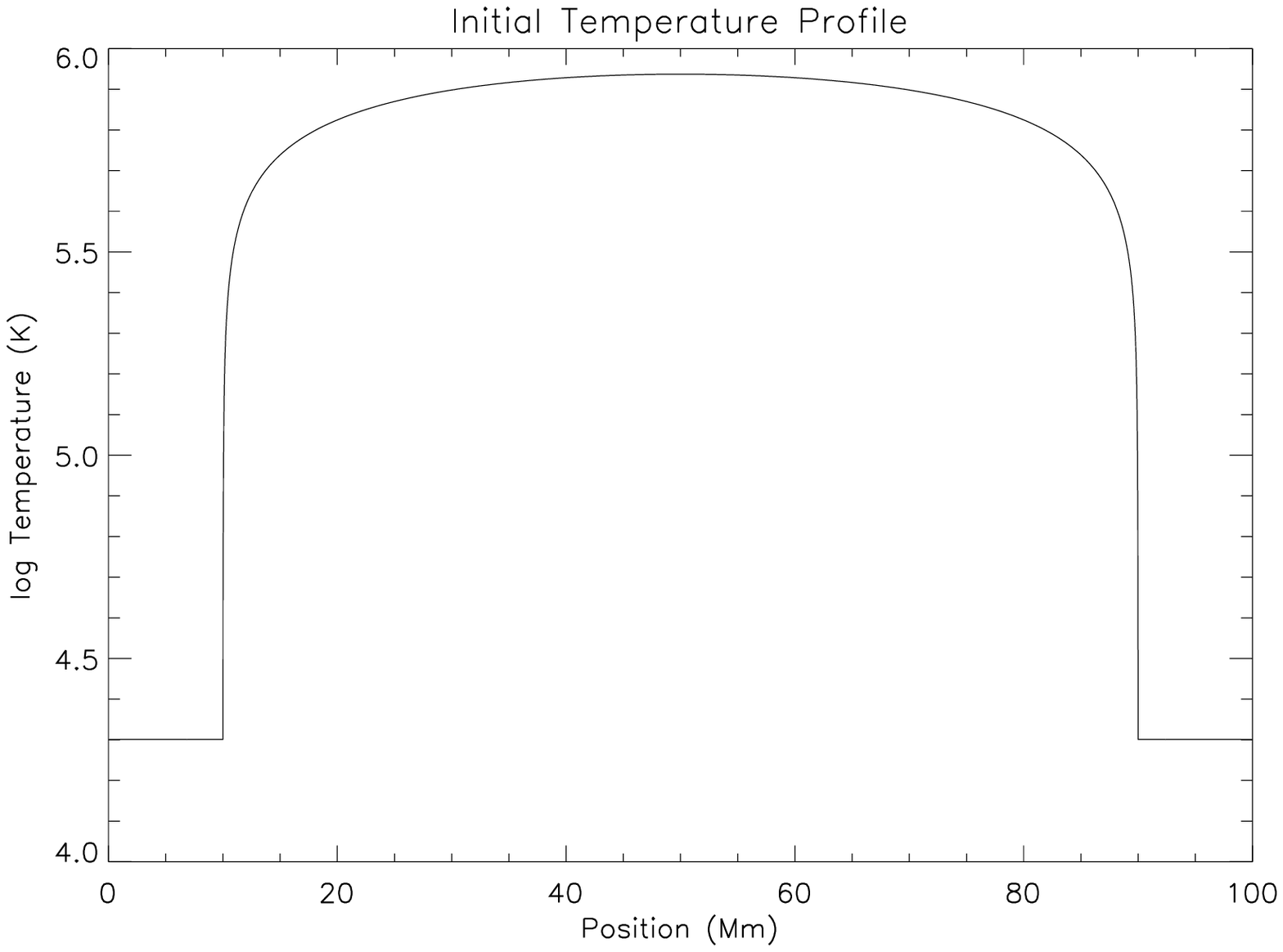}
\includegraphics[width=0.8\textwidth]{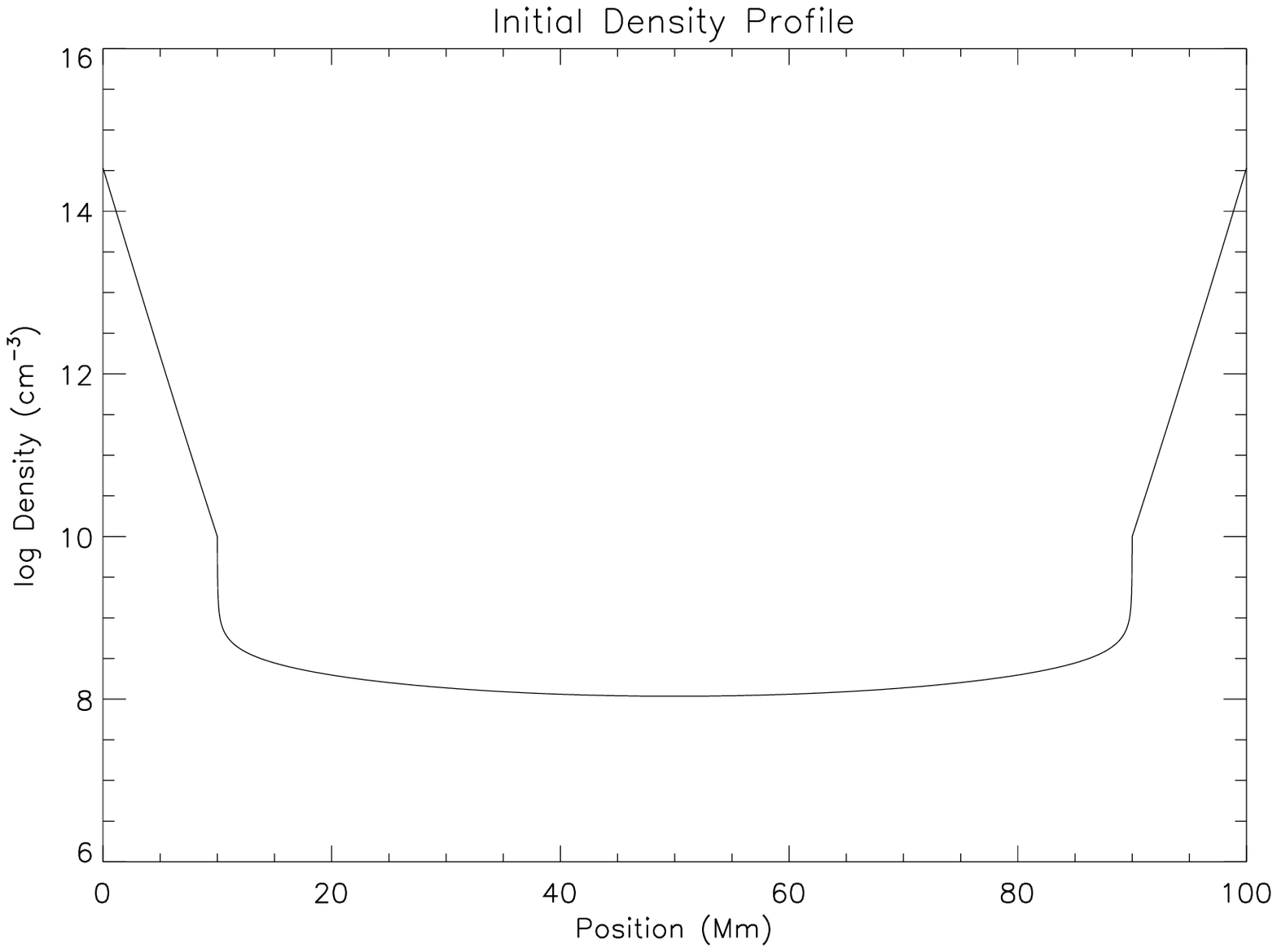}
\caption{The initial hydrostatic temperature and density profiles for the set of numerical experiments where $2L = 80$ Mm.}
\label{fig2}
\end{figure}

The loop geometry is semi-circular with a $2\times10^4$~K chromosphere of depth $10^{9}$~cm (several scale heights) at each footpoint, providing a reservoir of mass and energy that will be supplied to the corona upon heating.  Figure~\ref{fig2} shows the initial temperature and density profiles for the set of numerical experiments where $2L=80$~Mm, which were found by integrating the hydrostatic equations from foot-point to foot-point. The initial conditions are energetically negligible compared with the magnitude of energy released during the nanoflare train.

The forward modeling aspect of this work is described in detail by Bradshaw \& Klimchuk (2011) and in Paper~I, though we summarize the salient information here. A multi-stranded loop is created such that each strand represents a one second interval of the entire heating and cooling cycle. For example, if a heating and cooling cycle took 3600~s (1 hour) then our model loop would consist of 3600 individual strands \citep[e.g.][]{testa2005,guarrasi2010}. Since each of the strands evolve independently we believe that this procedure provides a reasonable representation of the average state of the loop atmosphere when observed. The EM is then calculated in two separate ways. We first calculate an EM that has no dependence on instrumental or line-of-sight constraints by directly evaluating the integral ($EM_{C} = \int n^{2} dr$).  This is the true, or model, EM. We can do this because we have access to the values of density and temperature for each strand, calculated by our numerical model, a luxury not available to observers. We calculate the apex EM (to avoid foot-point~/~moss contamination) and sum over all of the strands to find the total EM.

\begin{table}
\caption{The 30 spectral lines used to compute the synthetic Hinode-EIS emission measure.}
\centering
\begin{tabular}{c c}
\centering
\begin{tabular}{c c c}
\tableline
Ion & Wavelength & $\log_{10}~T$ \\ \tableline
Mg V & 276.579 & 5.45 \\
Mg VI & 268.991 & 5.65 \\
Mg VI & 270.391 &  5.65 \\
Si VII & 275.354 & 5.80 \\
Mg VII & 278.404 & 5.80 \\
Mg VII & 280.745 & 5.80 \\
Fe IX & 188.497 & 5.85 \\
Fe IX & 197.865 & 5.85 \\
Si IX & 258.082 & 6.05 \\
Fe X & 184.357 & 6.05 \\
Fe XI & 180.408 & 6.15 \\
Fe XI & 188.232 & 6.15 \\
Si X & 258.371 & 6.15 \\
Si X & 261.044 & 6.15 \\
S X & 264.231 & 6.15 \\
\end{tabular}
\hspace{0.1in}
\quad
\begin{tabular}{c c c}
\tableline
Ion & Wavelength & $\log_{10}~T$ \\ \tableline
Fe XII & 192.394 & 6.20 \\
Fe XII & 195.119 & 6.20 \\
Fe XIII & 202.044 & 6.25 \\
Fe XIII & 203.828 & 6.25 \\
Fe XIV & 264.790 & 6.30 \\
Fe XIV & 270.522 & 6.30 \\
Fe XIV & 274.204 & 6.30 \\
Fe XV & 284.163 & 6.35 \\
S XIII & 256.685 & 6.40 \\
Fe XVI & 262.976 & 6.45 \\
Ca XIV & 193.866 & 6.55 \\
Ca XV & 200.972 & 6.65 \\
Ca XVI & 208.604 & 6.70 \\
Ca XVII & 192.853 & 6.75 \\
Fe XVII & 269.494 & 6.75 \\
\end{tabular}
\end{tabular}
\label{table2}
\end{table}

We also calculate a synthetic EM along the line-of-sight that Hinode-EIS would see if it were to observe loops in an active region that were identical to our model loops. We use the Pottasch method \citep{pottasch1963,jordan1987,tripathi2011} and construct an apex EM for each strand (summed to find the total EM) from 30 spectral lines chosen from published observational studies. These lines are formed across a wide range of temperatures and are listed in Table~\ref{table2}. The forward-modeling procedure by which the line intensities were synthesized using the concept of a virtual detector is described in detail in Section~3 of \cite{bradshaw2011} and summarized in Appendix~\ref{appA}. A set of EM-loci curves can then be derived by dividing each line intensity by the contribution function \citep[calculated from atomic data provided by the Chianti database:][]{dere1997,dere2009}. Following \cite{pottasch1963} the values for the EM assigned to each line are estimated by assuming that the contribution function is a square function having a constant value over a temperature range of width $\Delta \log T = 0.3$. We have used a density of $10^9$~cm$^{-3}$ in the contribution functions because we showed in Paper~I that this value brings density sensitive lines into better agreement with their neighbors.

The EM derived in this manner is subject to some of the same constraints as a real observed loop and can be compared to published EMs derived from observational studies. By calculating these two EMs, we can see what features of the true EM are reliably reproduced in the synthetic one and what information about the state of the plasma is lost. We can then establish a level of confidence in the information that we extract from an observed EM. \cite{testa2012} have carried out an extensive study of this important issue by comparing known model EMs from a 3D model of a quiet Sun region, with EMs derived by tracing different lines-of-sight through the computational domain and treating the summed emission as though it were real data. They used the Extreme-ultraviolet Imaging Spectrometer (Hinode-EIS) response functions to calculate synthetic spectra and the Atmospheric Imaging Assembly (SDO-AIA) response functions to calculate synthetic intensities for its wavelength channels. The EMs were then constructed using the Monte Carlo Markov Chain (MCMC) method. They found that the EMs derived from EIS synthetic data were able to reproduce some characteristics of the model distributions, but showed differences when structures with significantly different density intersected the line-of-sight. The EMs derived from AIA synthetic data were much less accurate. We calculate synthetic EIS spectra in a similar manner as \cite{testa2012}, by taking lines-of-sight through the apex of our target loop. We are primarily interested in AR cores and if these can be observed directly from above at disk center then it is reasonable to suppose that they are the densest structures along the line-of-sight, and therefore should determine the magnitude of the EM. However, we appreciate that this is an important observational issue that deserves attention if EMs are to
be used as a diagnostic of timescale related to coronal heating and we base our estimates of the uncertainties associated with the EM slope derived from observations on the recent work of
\cite{guennou2012a,guennou2012b,guennou2013}.

\section{Results}
\label{results}

\begin{longtable}{c c c c c c c c c c c}
\caption{The results of 60 numerical experiments for heating by nanoflare trains.} \\
\hline
Run \# & $2L$ & $N$ & $\tau_{H}$ & $\tau_{C}$ & $\Delta_{H}$ & $\Delta_{C}$ & $\frac{\Delta_{H}}{\Delta_{C}}$ & log$_{10}$ $T_{\mbox{peak}}$ & $\alpha_{\mbox{model}}$ & $\alpha_{\mbox{observed}}$\\
\hline
\endhead
1 & 40 & 5 & 60 & 60 & 540 & 2203 & .245 & 6.45 & 1.07 ($\pm$ 0.10) & 0.91 ($\pm$ 0.20) \\
2 & 40 & 10 & 60 & 60 & 1140 & 1949 & .585 & 6.55 & 1.61 ($\pm$ 0.11) & 1.55 ($\pm$ 0.29) \\
3 & 40 & 15 & 60 & 60 & 1740 & 1866 & .932 & 6.55 & 1.94 ($\pm$ 0.28) & 1.93 ($\pm$ 0.35) \\
4 & 40 & 20 & 60 & 60 & 2340 & 1823 & 1.28 & 6.55 & 2.18 ($\pm$ 0.40)& 2.24 ($\pm$ 0.40) \\
5 & 40 & 5 & 60 & 300 & 1500 & 4198 & .357 & 6.35 & 1.92 ($\pm$ 0.34) & 1.61 ($\pm$ 0.25) \\
6 & 40 & 10 & 60 & 300 & 3300 & 3844 & .858 & 6.35 & 2.78 ($\pm$ 0.82) & 2.93 ($\pm$ 0.36) \\
7 & 40 & 15 & 60 & 300 & 5100 & 3776 & 1.35 & 6.35 & 3.30 ($\pm$ 1.08) & 3.74 ($\pm$ 0.41) \\
8 & 40 & 20 & 60 & 300 & 6900 & 3771 & 1.83 & 6.35 & 3.59 ($\pm$ 1.26)& 4.26 ($\pm$ 0.44) \\
9 & 40 & 5 & 180 & 180 & 1620 & 3767 & .430 & 6.55 & 1.90 ($\pm$ 0.18) & 1.98 ($\pm$ 0.35) \\
10 & 40 & 10 & 180 & 180 & 3420  & 3453 & .990 & 6.55 & 2.49 ($\pm$ 0.42) & 2.76 ($\pm$ 0.47) \\
11 & 40 & 15 & 180 & 180 & 5220 & 3429 & 1.52 & 6.55 & 2.84 ($\pm$ 0.52) & 3.28 ($\pm$ 0.54) \\
12 & 40 & 20 & 180 & 180 & 7020 & 3436 & 2.04 & 6.55 & 3.07 ($\pm$ 0.59) & 3.65 ($\pm$ 0.59) \\
13 & 40 & 5 & 300 & 60 & 1740 & 3453 & .504 & 6.65 & 1.99 ($\pm$ 0.16) & 1.68 ($\pm$ 0.20) \\
14 & 40 & 10 & 300 & 60 & 3540 & 3292 & 1.08 & 6.65 & 2.54 ($\pm$ 0.33) & 2.41 ($\pm$ 0.25) \\
15 & 40 & 15 & 300 & 60 & 5340 & 3270 & 1.63 & 6.65 & 2.87 ($\pm$ 0.41) & 2.81 ($\pm$ 0.28) \\
16 & 40 & 20 & 300 & 60 & 7140 & 3284 & 2.17 & 6.65 & 3.06 ($\pm$ 0.49) & 3.07 ($\pm$ 0.31) \\
17 & 40 & 5 & 300 & 300 & 2700 & 3530 & .765 & 6.55 & 2.43 ($\pm$ 0.30) & 2.64 ($\pm$ 0.43) \\
18 & 40 & 10 & 300 & 300 & 5700 & 3448 &  1.65 & 6.55 & 3.08 ($\pm$ 0.52) & 3.59 ($\pm$ 0.54) \\
19 & 40 & 15 & 300 & 300 & 8700 & 3446 & 2.52 & 6.55 & 3.42 ($\pm$ 0.63) & 4.15 ($\pm$ 0.59) \\
20 & 40 & 20 & 300 & 300 & 11700 & 3457 & 3.38 & 6.55 & 3.65 ($\pm$ 0.72) & 4.56 ($\pm$ 0.63) \\
21 & 80 & 5 & 60 & 60 & 540 & 4333 & .125 & 6.35 & 1.23 ($\pm$ 0.06) & 0.88 ($\pm$ 0.12) \\
22 & 80 & 10 & 60 & 60 & 1140 & 3939 & .289 & 6.65 & 1.11 ($\pm$ 0.12) & 1.49 ($\pm$ 0.17) \\
23 & 80 & 15 & 60 & 60 & 1740 & 3699 & .470 & 6.65 & 1.53 ($\pm$ 0.07) & 1.77 ($\pm$ 0.22) \\
24 & 80 & 20 & 60 & 60 & 2340 & 3538 & .661 & 6.65 & 1.77 ($\pm$ 0.14) & 1.97 ($\pm$ 0.26) \\
25 & 80 & 5 & 60 & 300 & 1500 & 4198 & .357 & 6.45 & 1.29 ($\pm$ 0.06) & 1.15 ($\pm$ 0.13) \\
26 & 80 & 10 & 60 & 300 & 3300 & 3844 & .858 & 6.45 & 2.16 ($\pm$ 0.36) & 2.10 ($\pm$ 0.22) \\
27 & 80 & 15 & 60 & 300 & 5100 & 3776 & 1.35 & 6.45 & 2.57 ($\pm$ 0.59) & 2.76 ($\pm$ 0.30) \\
28 & 80 & 20 & 60 & 300 & 6900 & 3771 & 1.83 & 6.45 & 2.84 ($\pm$ 0.73) & 3.26 ($\pm$ 0.35) \\
29 & 80 & 5 & 180 & 180 & 1620 & 3767 & .430 & 6.55 & 1.61 ($\pm$ 0.04)& 1.81 ($\pm$ 0.22) \\
30 & 80 & 10 & 180 & 180 & 3420  & 3453 & .990 & 6.65 & 2.15 ($\pm$ 0.16) & 2.24 ($\pm$ 0.21) \\
31 & 80 & 15 & 180 & 180 & 5220 & 3429 & 1.52 & 6.65 & 2.49 ($\pm$ 0.28) & 2.69 ($\pm$ 0.26) \\
32 & 80 & 20 & 180 & 180 & 7020 & 3436 & 2.04 & 6.65 & 2.72 ($\pm$ 0.35) & 3.00 ($\pm$ 0.30) \\
33 & 80 & 5 & 300 & 60 & 1740 & 3453 & .504 & 6.65 & 1.66 ($\pm$ 0.07) & 1.38 ($\pm$ 0.13) \\
34 & 80 & 10 & 300 & 60 & 3540 & 3292 & 1.08 & 6.65 & 2.09 ($\pm$ 0.23) & 2.03 ($\pm$ 0.15) \\
35 & 80 & 15 & 300 & 60 & 5340 & 3270 & 1.63 & 6.65 & 2.30 ($\pm$ 0.35) & 2.40 ($\pm$ 0.17) \\
36 & 80 & 20 & 300 & 60 & 7140 & 3284 & 2.17 & 6.65 & 2.44 ($\pm$ 0.43) & 2.65 ($\pm$ 0.19) \\
37 & 80 & 5 & 300 & 300 & 2700 & 3530 & .765 & 6.55 & 1.99 ($\pm$ 0.18) & 2.20 ($\pm$ 0.30) \\
38 & 80 & 10 & 300 & 300 & 5700 & 3448 &  1.65 & 6.65 & 2.57 ($\pm$ 0.31) & 2.82 ($\pm$ 0.28) \\
39 & 80 & 15 & 300 & 300 & 8700 & 3446 & 2.52 & 6.65 & 2.88 ($\pm$ 0.42) & 3.26 ($\pm$ 0.33) \\
40 & 80 & 20 & 300 & 300 & 11700 & 3457 & 3.38 & 6.65 & 3.07 ($\pm$ 0.49) & 3.57 ($\pm$ 0.37) \\
41 & 160 & 5 & 60 & 60 & 540 & 6612 & .082 & 6.35 & 0.79 ($\pm$ 0.07) & 1.18 ($\pm$ 0.65) \\
42 & 160 & 10 & 60 & 60 & 1140 & 6585 & .173 & 6.45 & 1.27 ($\pm$ 0.14) & 1.27 ($\pm$ 0.21) \\
43 & 160 & 15 & 60 & 60 & 1740 & 6577 & .265 & 6.65 & 1.09 ($\pm$ 0.15) & 1.61 ($\pm$ 0.21) \\
44 & 160 & 20 & 60 & 60 & 2340 & 6579 & .356 & 6.65 & 1.41 ($\pm$ 0.11) & 1.82 ($\pm$ 0.21) \\
45 & 160 & 5 & 60 & 300 & 1500 & 6006 & .250 & 6.35 & 0.96 ($\pm$ 0.07) & 1.22 ($\pm$ 0.67) \\
46 & 160 & 10 & 60 & 300 & 3300 & 6099 & .541 & 6.45 & 1.62 ($\pm$ 0.11) & 1.50 ($\pm$ 0.23) \\
47 & 160 & 15 & 60 & 300 & 5100 & 6112 & .834 & 6.45 & 2.05 ($\pm$ 0.29) & 2.04 ($\pm$ 0.26) \\
48 & 160 & 20 & 60 & 300 & 6900 & 6082 & 1.13 & 6.45 & 2.35 ($\pm$ 0.43) & 2.46 ($\pm$ 0.29) \\
49 & 160 & 5 & 180 & 180 & 1620 & 6718 & .241 & 6.55 & 1.32 ($\pm$ 0.14) & 1.63 ($\pm$ 0.21) \\
50 & 160 & 10 & 180 & 180 & 3420 & 6567 & .521 & 6.65 & 1.78 ($\pm$ 0.06) & 2.11 ($\pm$ 0.24) \\
51 & 160 & 15 & 180 & 180 & 5220 & 6434 & .811 & 6.65 & 2.12 ($\pm$ 0.15) & 2.25 ($\pm$ 0.19) \\
52 & 160 & 20 & 180 & 180 & 7020 & 6425 & 1.09 & 6.65 & 2.35 ($\pm$ 0.21) & 2.55 ($\pm$ 0.22) \\
53 & 160 & 5 & 300 & 60 & 1740 & 6762 & .257 & 6.65 & 1.37 ($\pm$ 0.14) & 1.94 ($\pm$ 0.21) \\
54 & 160 & 10 & 300 & 60 & 3540 & 6463 & .548 & 6.65 & 1.86 ($\pm$ 0.07) & 1.64 ($\pm$ 0.13) \\
55 & 160 & 15 & 300 & 60 & 5340 & 6344 & .842 & 6.65 & 2.13 ($\pm$ 0.15) & 2.02 ($\pm$ 0.13) \\
56 & 160 & 20 & 300 & 60 & 7140 & 6282 & 1.14 & 6.65 & 2.30 ($\pm$ 0.24) & 2.28 ($\pm$ 0.15) \\
57 & 160 & 5 & 300 & 300 & 2700 & 6606 & .409 & 6.65 & 1.55 ($\pm$ 0.09) & 2.02 ($\pm$ 0.23) \\
58 & 160 & 10 & 300 & 300 & 5700 & 6530 & .873 & 6.65 & 2.22 ($\pm$ 0.14) & 2.38 ($\pm$ 0.20) \\
59 & 160 & 15 & 300 & 300 & 8700 & 6471 & 1.34 & 6.65 & 2.56 ($\pm$ 0.25) & 2.81 ($\pm$ 0.25) \\
60 & 160 & 20 & 300 & 300 & 11700 & 6490 & 1.80 & 6.65 & 2.79 ($\pm$ 0.32) & 3.12 ($\pm$ 0.28) \\
\label{table3}
\end{longtable}

\begin{figure}
\begin{minipage}[b]{0.5\linewidth}
\centering
\includegraphics[width=2.95in]{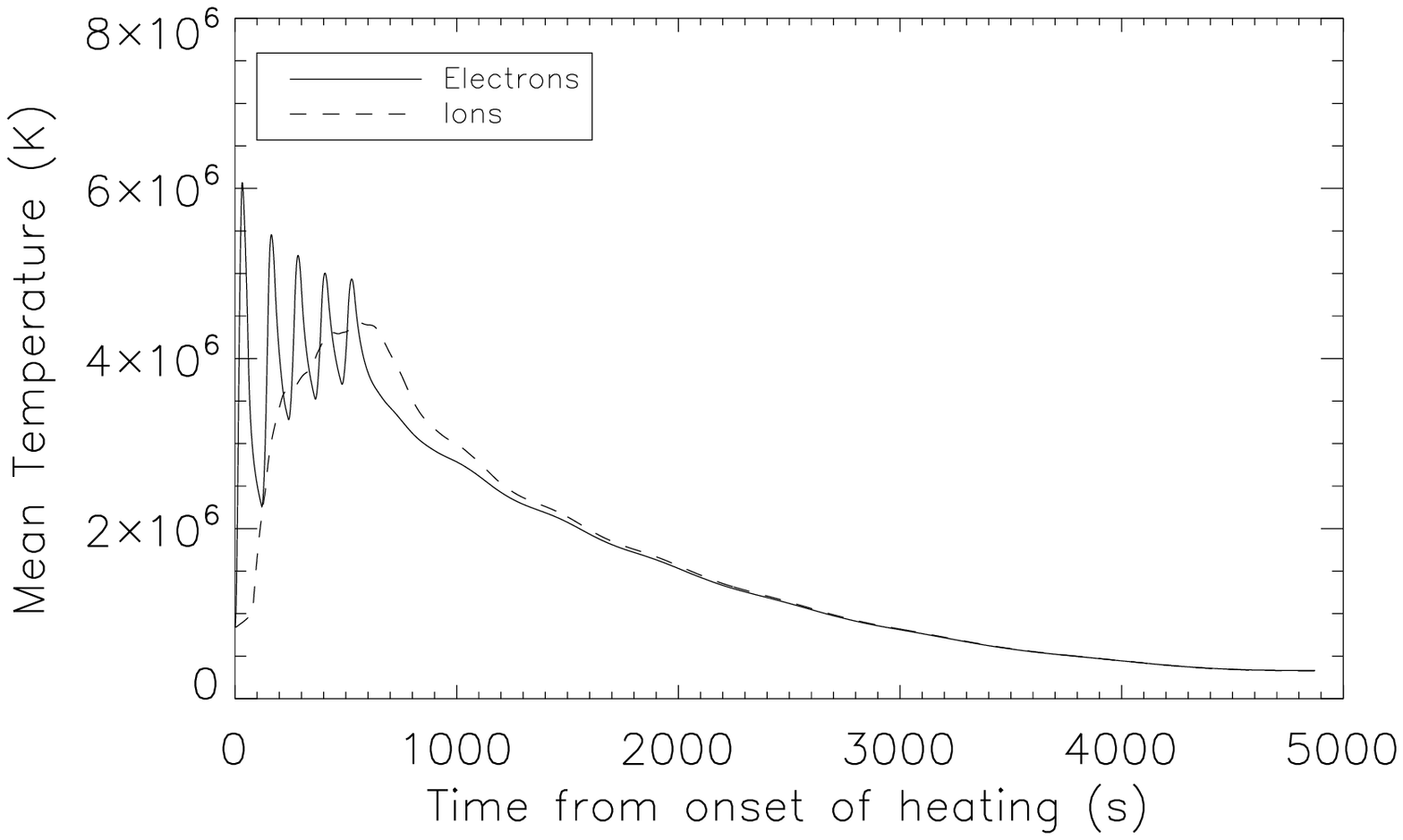}
\end{minipage}
\hspace{0.1in}
\begin{minipage}[b]{0.5\linewidth}
\centering
\includegraphics[width=2.95in]{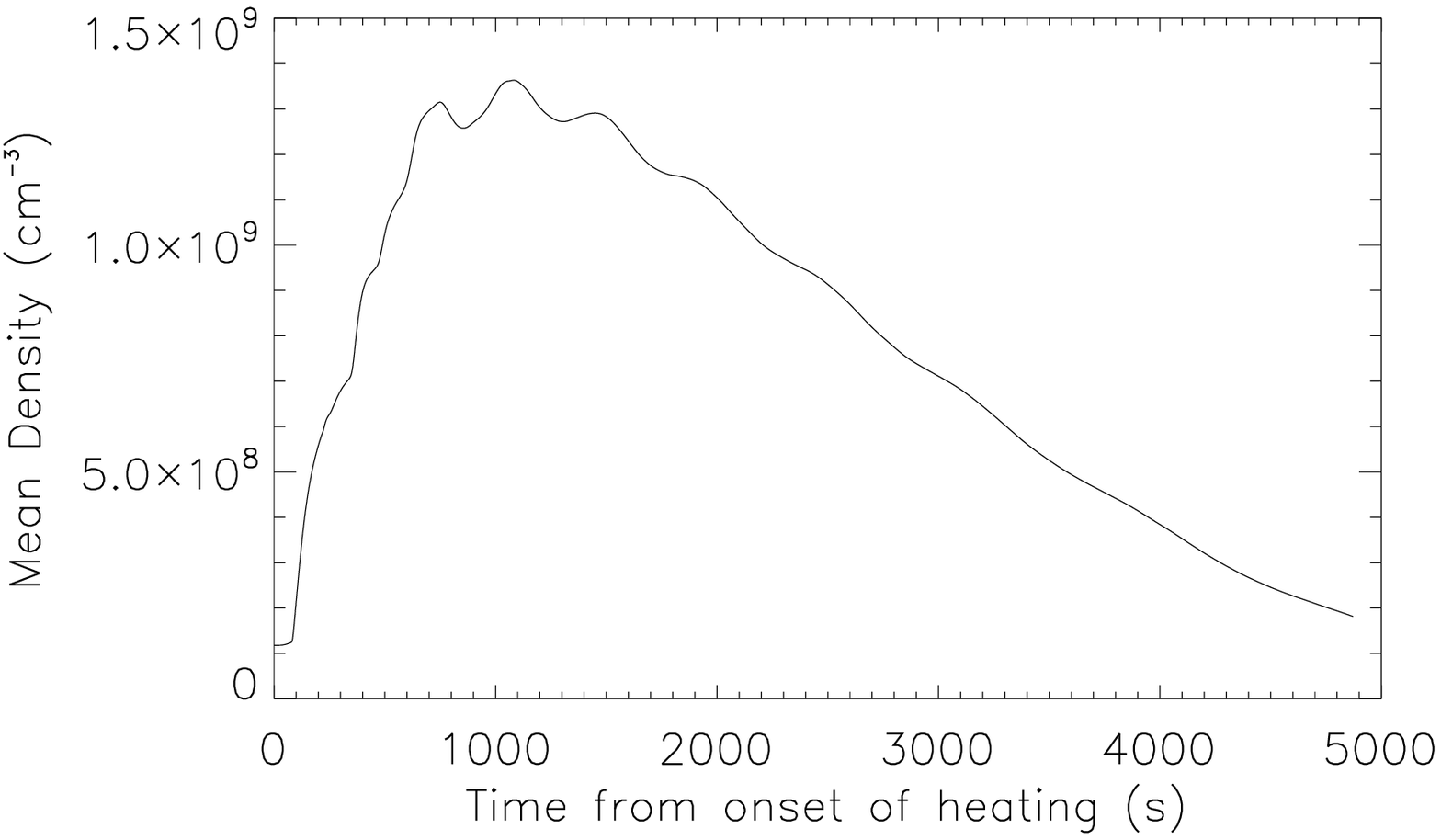}
\end{minipage}
\includegraphics[width=6.00in]{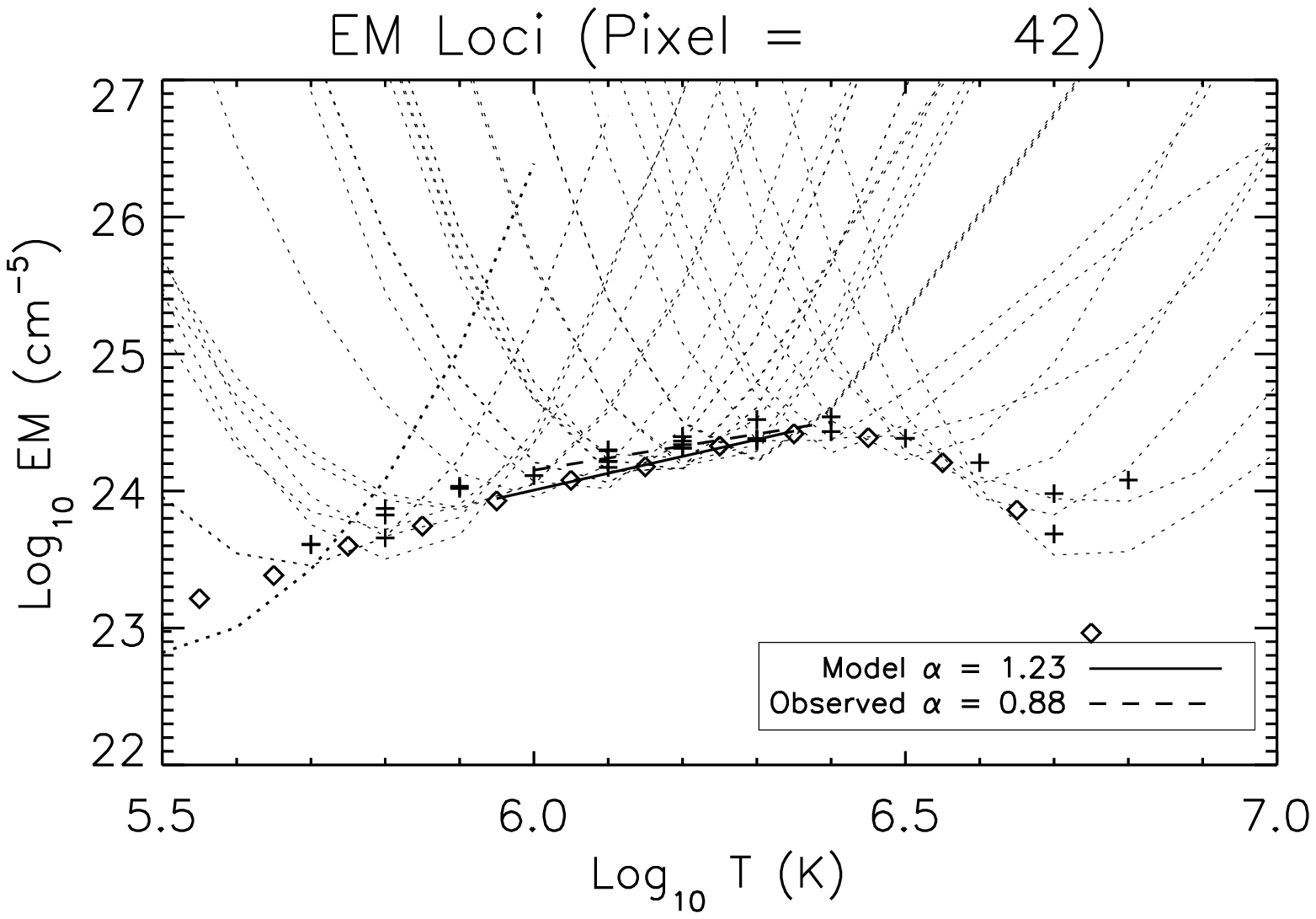}
\caption{Results for Run~21. Upper-left panel: time-evolution of the average electron and ion temperatures. Upper-right panel: time-evolution of the average coronal electron density. Averages are calculated over the upper 10\% of the loop, centered on the apex. Lower-panel: model emission measure (diamonds) and the emission measure derived from synthetic Hinode-EIS data (plus-signs). Pixel 42 indicates the apex pixel on the model detector.}
\label{fig3}
\end{figure}

We performed 60 numerical experiments for nanoflare train heating within the parameter space described in Section~\ref{modeling}. The results of these experiments are summarized in Table~\ref{table3}. As an example consider Run~21, for which the temporal profile of the enegy release by a nanoflare train is shown in Figure~\ref{fig1}. The nanoflare train consists of $N=5$ separate heating events of period $\tau_H = 60$~s each, with an interval of $\tau_C = 60$~s between each event. The total heating time $\Delta_H$ is the time from the onset of the first heating event until the cessation of the final heating event. The total cooling time $\Delta_C$ is the time from the end of the final heating event until the electron temperature falls below $10^5$~K. The temporal evolution of the average coronal electron and ion temperatures, and the average electron density, are shown in the upper two panels of Figure~\ref{fig3}. The model and synthetic observed EMs derived from the results of Run~21 are shown in the lower panel of Figure~\ref{fig3}, where the diamonds are the model values and the plus-signs are the synthetic quantities derived using the Pottasch method. EM loci curves are shown as dotted lines. The EM peaks at a temperature $T_{\mbox{peak}}$ in the region of $10^{6.35}$~K. A linear regression applied to the model and synthetic quantities between $T_{\mbox{peak}}$ and $10^6$~K yields slopes of $\alpha_{\mbox{model}}=1.23$ and
$\alpha_{\mbox{observed}}=0.88$. The quantity in brackets that accompanies each slope in Table~\ref{table3} is the one-sigma uncertainty estimate of the slope (found with the standard IDL procedure LINFIT). It is important to remember that the same atomic physics quantities are used to infer the EM from the synthetic line intensities as are used to derive the intensities from the model.  In the case of actual observations, the adopted atomic physics quantities are likely to be different from the true quantities, so the uncertainties in the slope are much larger \citep{guennou2012a,guennou2012b,guennou2013}.

Based on the results presented in Table~\ref{table3}, we can make a number of observations concerning the relationship between the parameters that we have explored and the EM slopes that we have found. For each set of 20 runs, where $2L$ remains fixed, it can be seen that steeper slopes are obtained with increasing $N$ when $\tau_H$ and $\tau_C$ are fixed. In the case of Runs~[8,20] for $N=20$ we find $\alpha_{\mbox{observed}}>4$, which is significantly steeper than any of the EM slopes we found in our previous investigation of low-frequency nanoflares. This indicates, as expected from the discussion in Section~\ref{introduction}, that sustaining the emission close to the temperature of peak emission measure for an extended period is an important element of obtaining slopes that are comparable to the upper-range of those that have been observed.

We also find that when considering pairs of Runs where only $\tau_H$ varies (e.g. [1,13], [2,14],...,[8,20]), the slope is always steeper for the Run with longer $\tau_H$, with no exceptions. Nonetheless, longer heating timescales for the individual nanoflares of the train are not sufficient by themselves to guarantee steeper EM slopes. Consider the group of Runs~[4,8,12,16,20] for all of which $N=20$. Run~4 has the shallowest slope and Run~20 the steepest slope, which may fit with expectations. However, Run~8 has a significantly steeper slope than both Runs~12 and 16. It is clear that adjusting $\tau_H$ by itself is not sufficient to guarantee slopes towards the upper-range of those observed and in consequence we may conclude that the slope of the EM is not sufficient by itself to constrain $\tau_H$.

Further examination of Table~\ref{table3} yields a connection between the time interval $\tau_C$ before the next nanoflare of the train and the EM slope. If we consider pairs of Runs for which $N$ and $\tau_H$ are fixed then a pattern emerges. Runs~[4,8] and [16,20] both show steeper slopes for longer $\tau_C$. The pattern persists when we include Runs~[9-12] in our analysis, which have intermediate heating timescales (e.g compare Runs [8,12] and [12,16]). Longer time intervals between successive nanoflares consistently yield steeper slopes. This relationship between $\tau_C$ and $\alpha_{\mbox{observed}}$ suggests that while sustaining the emission close to the temperature of peak emission measure via larger $N$ and $\tau_H$ is generally necessary for obtaining steeper slopes, the interval $\tau_C$ between individual nanoflares is also an important parameter when it is shorter than the characteristic cooling timescale of the plasma. This may partly be explained by comparing Runs~[8,16], [28,36] and [48,56], where each pair of Runs has $N=20$, $\tau_H=[60,300]~s$ and $\tau_C=[300,60]~s$. The Run with the shorter $\tau_H$ and longer $\tau_C$ of the pair always has the steeper slope but also a lower $\log_{10} T_{\mbox{peak}}$ and smaller $\Delta \log_{10} T$ between the EM peak and 1~MK. This may have the effect of steepening the slope, because the gradient is inversely proportional to $\Delta \log_{10} T$. We also expect increasing $\tau_C$ to effectively smear out the EM in temperature, as found by \cite{susino2010} who studied the effects of both uniform and localized heating in steady and impulsive regimes. They found that extended $\tau_C$ ($250-2000$~s) produced large oscillations in the coronal temperature and significant smearing of the EM across a broad temperature range, but we confine ourselves to generally shorter timescales for $\tau_C$ ($\le 300$~s) and expect any smearing of the EM to primarily affect the hot component. We discuss a method by which $\tau_C$ might be diagnosed from this effect toward the end of this Section.

\begin{figure}
\centering
\includegraphics{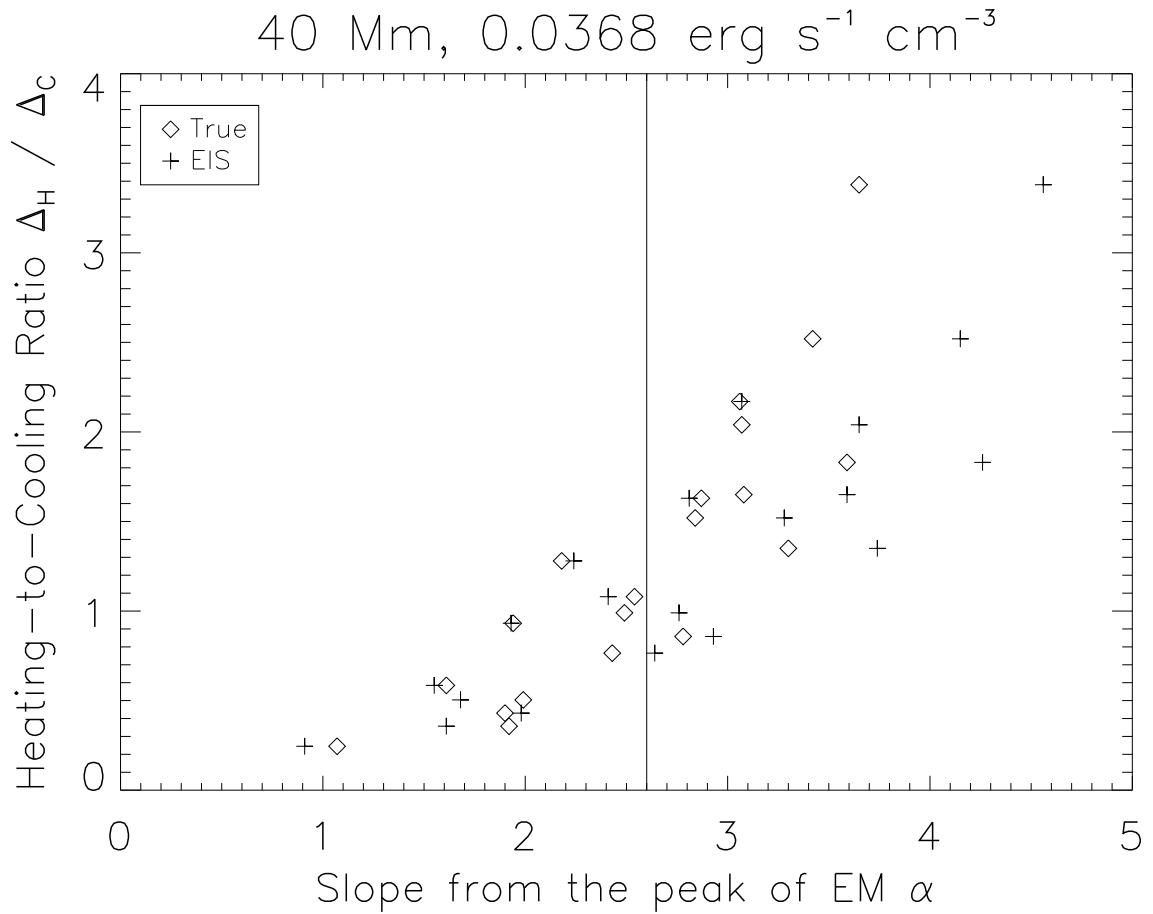}
\caption{The ratio $\Delta_H/\Delta_C$ versus $\alpha_{\mbox{model}}$ and $\alpha_{\mbox{observed}}$ for $2L=40$~Mm. The diamond-signs plot the true (model) emission measure calculations and the plus-signs plot the emission measure derived from synthetic Hinode-EIS data. The vertical line indicates the upper-limit to $\alpha_{\mbox{observed}}$ that can be explained with low-frequency nanoflares.}
\label{fig4}
\end{figure}

The total duration of the nanoflare train is given by $\Delta_H=N\tau_H+(N-1)\tau_C$ and the cooling time $\Delta_C$ following cessation of the final nanoflare was measured from the
numerical results. The quantity $\Delta_H/\Delta_C$, which is the ratio of `hot' strands (e.g. those affected by the nanoflare train) to cooler strands (e.g. those undergoing final cooling and
draining), is useful for parameterizing the relationship between [$N$, $\tau_H$, $\tau_C$] and $\alpha_{\mbox{observed}}$. Figure~\ref{fig4} shows the set of $\Delta_H/\Delta_C$ plotted against the values of $\alpha$ calculated from the model and synthetic emission measures. The vertical line at $\alpha=2.6$ in the figure delineates the boundary between slopes that are consistent with low-frequency nanoflares ($\alpha \le 2.6$) and those that are not ($\alpha>2.6$). The general trend that emerges is for larger values of $\Delta_H/\Delta_C$ to yield steeper EM slopes. Figure~\ref{fig4} also shows that slopes shallower than 2.6 may be consistent with both low-frequency nanoflares and nanoflare trains with relatively low $N$. The case of low-frequency nanoflares is essentially the limit $N=1$. We can now understand why increasing $\tau_C$ leads to steeper EM slopes; longer $\tau_C$ increases $\Delta_H$ relative to $\Delta_C$. One further trend that emerges in Figure~\ref{fig4} is for an increasing discrepancy between $\alpha_{\mbox{model}}$ and $\alpha_{\mbox{observed}}$ as the slope steepens, where the slope of the EM derived from synthetic Hinode-EIS data increasingly over-estimates the slope of the model EM. Since we have followed a forward modeling procedure to calculate $\alpha_{\mbox{observed}}$ then this suggests that the slopes found from EMs derived from real observational data might be subject to the same bias that underlies this discrepancy, leading to an over-estimate of slopes at the upper-range.

\begin{figure}
\centering
\includegraphics[width=0.8\textwidth]{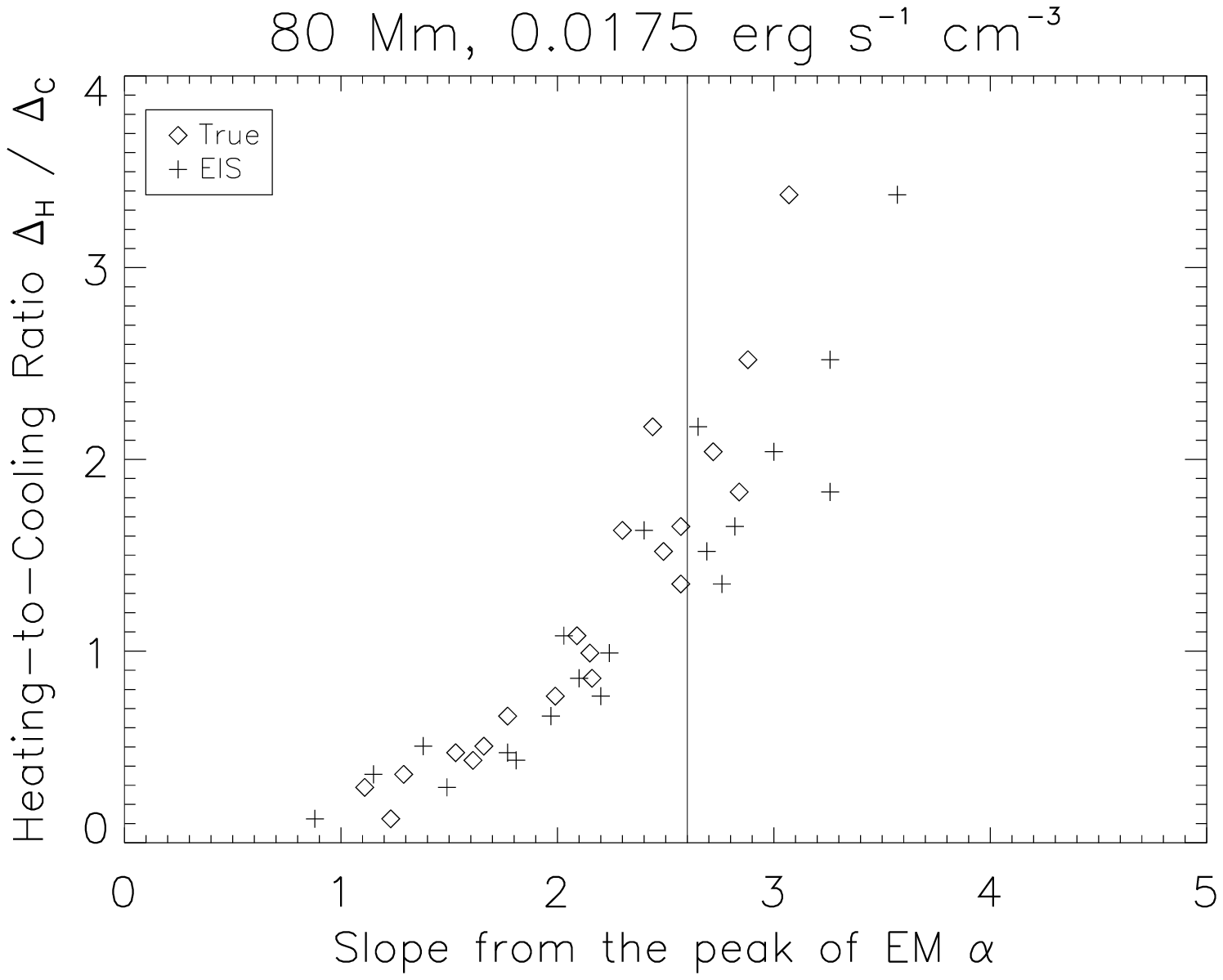}
\includegraphics[width=0.8\textwidth]{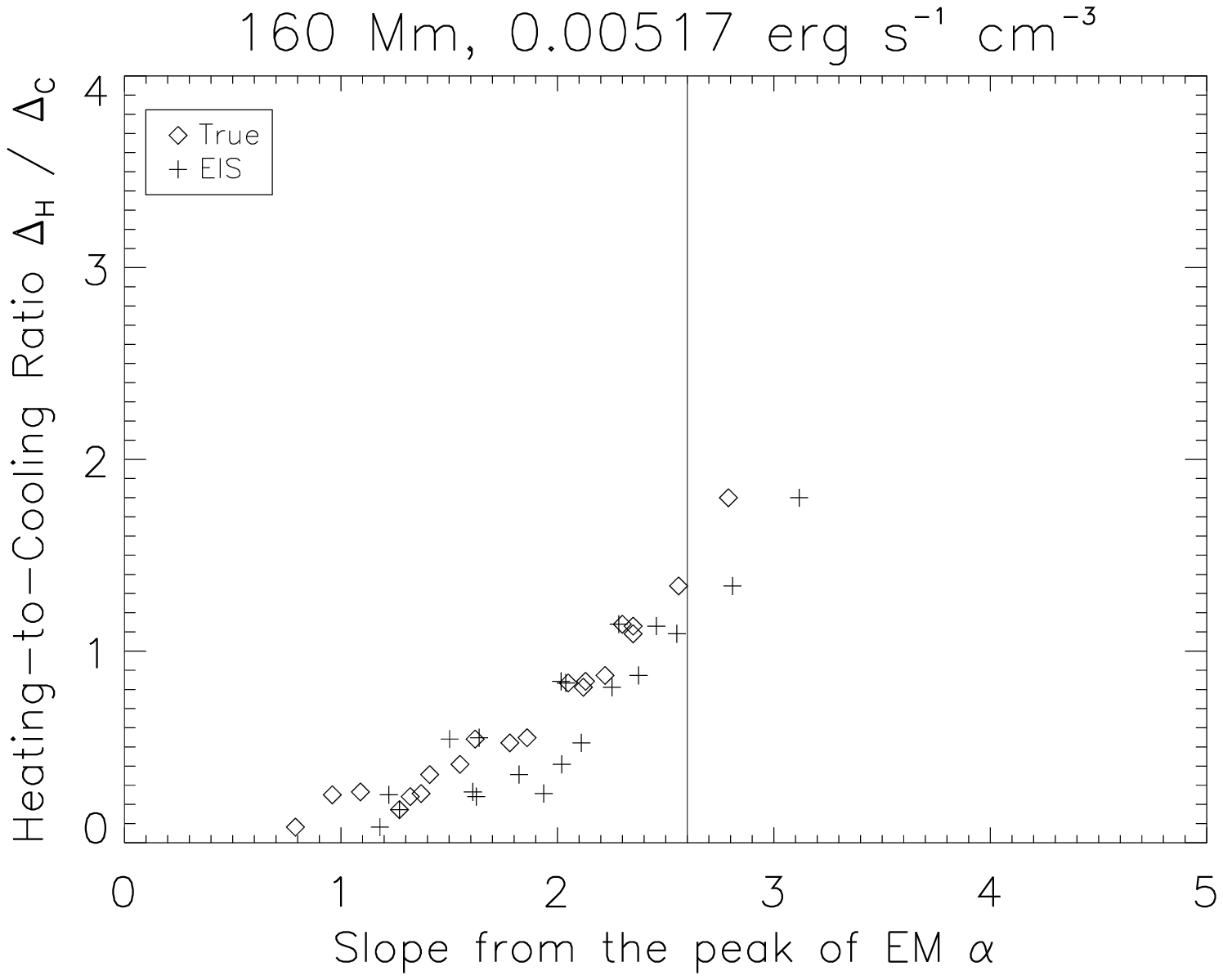}
\caption{The ratio $\Delta_H/\Delta_C$ versus $\alpha_{\mbox{model}}$ and $\alpha_{\mbox{observed}}$ for $2L=80$ and 160~Mm.}
\label{fig5}
\end{figure}

We now turn to considering the effect of changing $2L$ on the slopes obtained within our chosen nanoflare train parameter space. Comparing triplets of Runs for which $N$, $\tau_H$ and $\tau_C$ remain fixed and only $2L$ varies, e.g. Runs~[4,24,44], we find that $\alpha_{\mbox{observed}}$ is consistently shallower for longer loops. Figure~\ref{fig5} shows the $\Delta_H/\Delta_C$ versus $\alpha$ plots for $2L=[80,160]$~Mm where it becomes clear that the range of $\alpha$ is more restricted with increasing loop length. The steepest slopes for $2L=[80,160]$~Mm are $\alpha_{\mbox{observed}}=[3.57,3.12]$. Figure~\ref{fig5} also shows that an increasing proportion of the slopes yielded by nanoflare trains are consistent with the range of slopes yielded by low-frequency nanoflares. This is particularly evident for $2L=160$~Mm, where all but two values of $\alpha_{\mbox{observed}}$ are below 2.6. We can also see that the discrepancy between the slopes of the model EM and the EM from synthetic Hinode-EIS data persists for longer loops as the slope steepens. A comparison of the data points in Figures~\ref{fig4} and \ref{fig5} shows that this property does not appear related to $\Delta_H/\Delta_C$ because $\Delta_H/\Delta_C \approx 3.5$ in Figure~\ref{fig4} yields a larger difference between $\alpha_{\mbox{model}}$ and $\alpha_{\mbox{observed}}$ than $\Delta_H/\Delta_C \approx 3.5$ in the upper panel of Figure~\ref{fig5}.

We now discuss the physical reason for the dependence of the EM slope on the parameter $\Delta_H/\Delta_C$. Returning to Figure~\ref{fig3} for Run~21 ($N=5$, $\tau_H=60$~s, $\tau_C=60$~s) we see from the upper panels that the coronal density does not reach its peak until significantly after the nanoflare train has ended. Consequently, each strand of the loop is well into its final cooling phase at the time of peak EM and the data points from the temperature of peak EM to 1~MK are dominated by these cooling strands, which means that there is no significant contribution to the EM from the hot strands that are undergoing nanoflare train heating. The slope is therefore subject to the analytical limit $\alpha_{\mbox{max}}$ determined in Paper~I.

\begin{figure}
\begin{minipage}[b]{0.5\linewidth}
\centering
\includegraphics[width=2.95in]{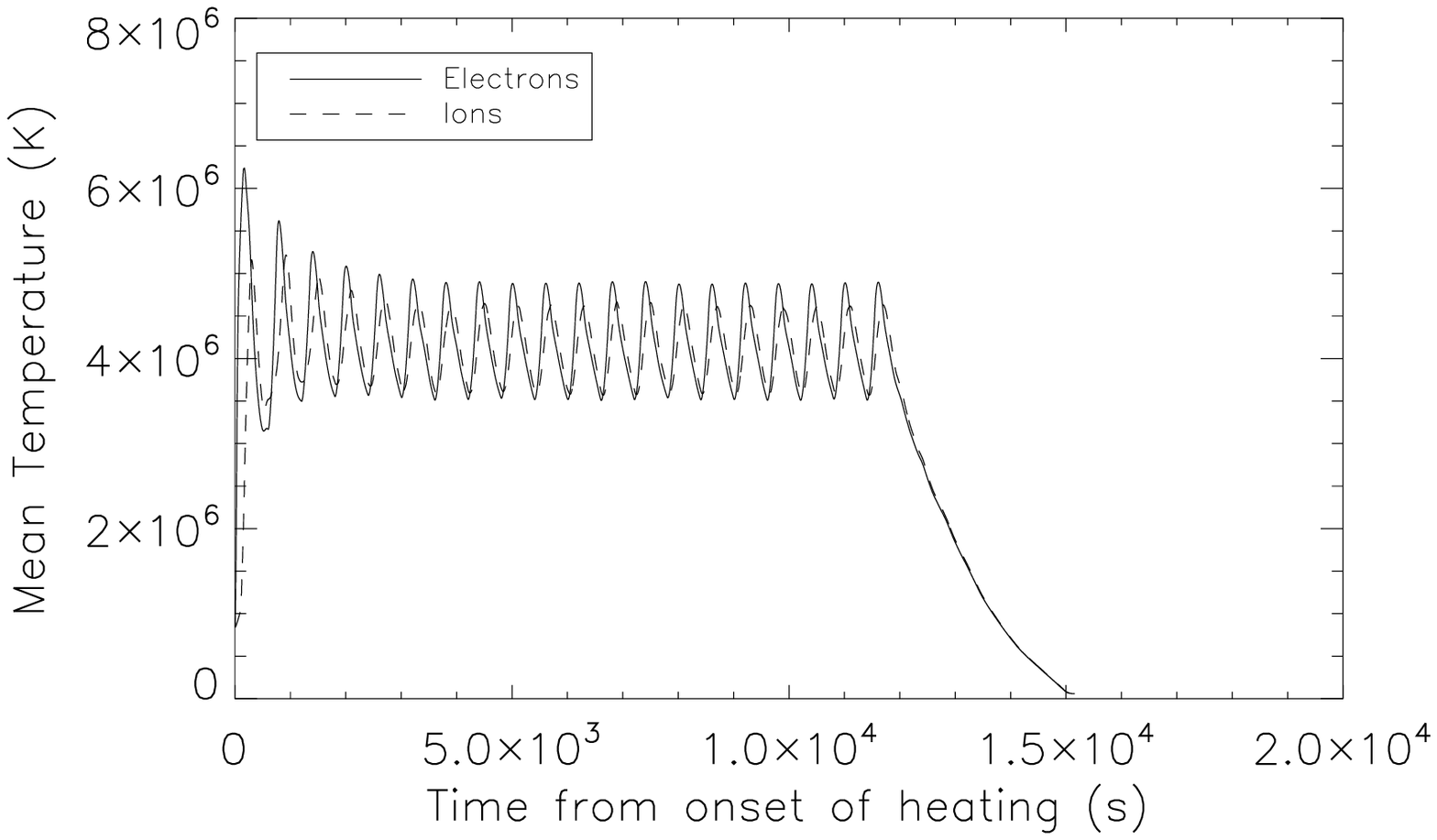}
\end{minipage}
\hspace{0.1in}
\begin{minipage}[b]{0.5\linewidth}
\centering
\includegraphics[width=2.95in]{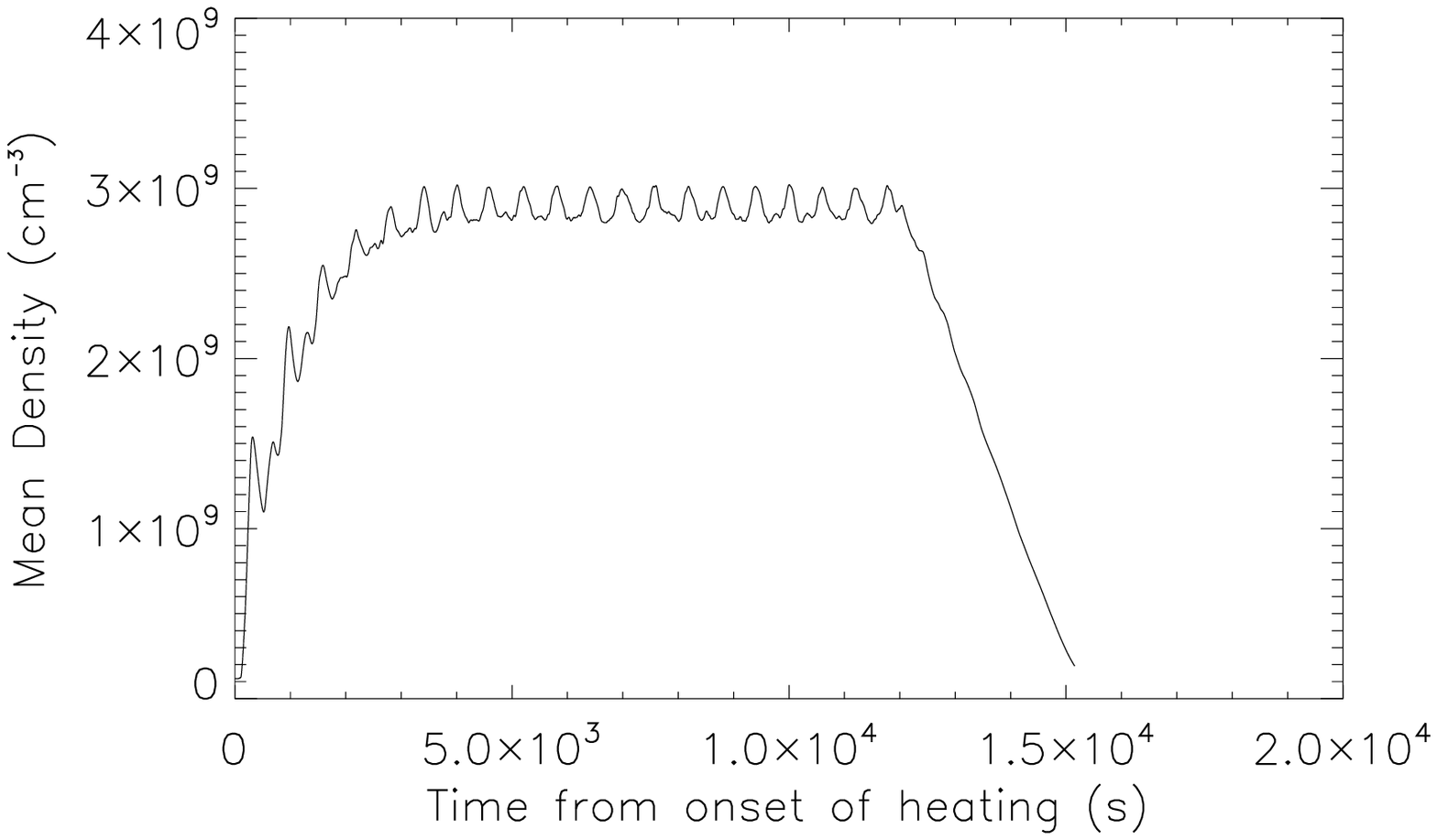}
\end{minipage}
\includegraphics[width=6.00in]{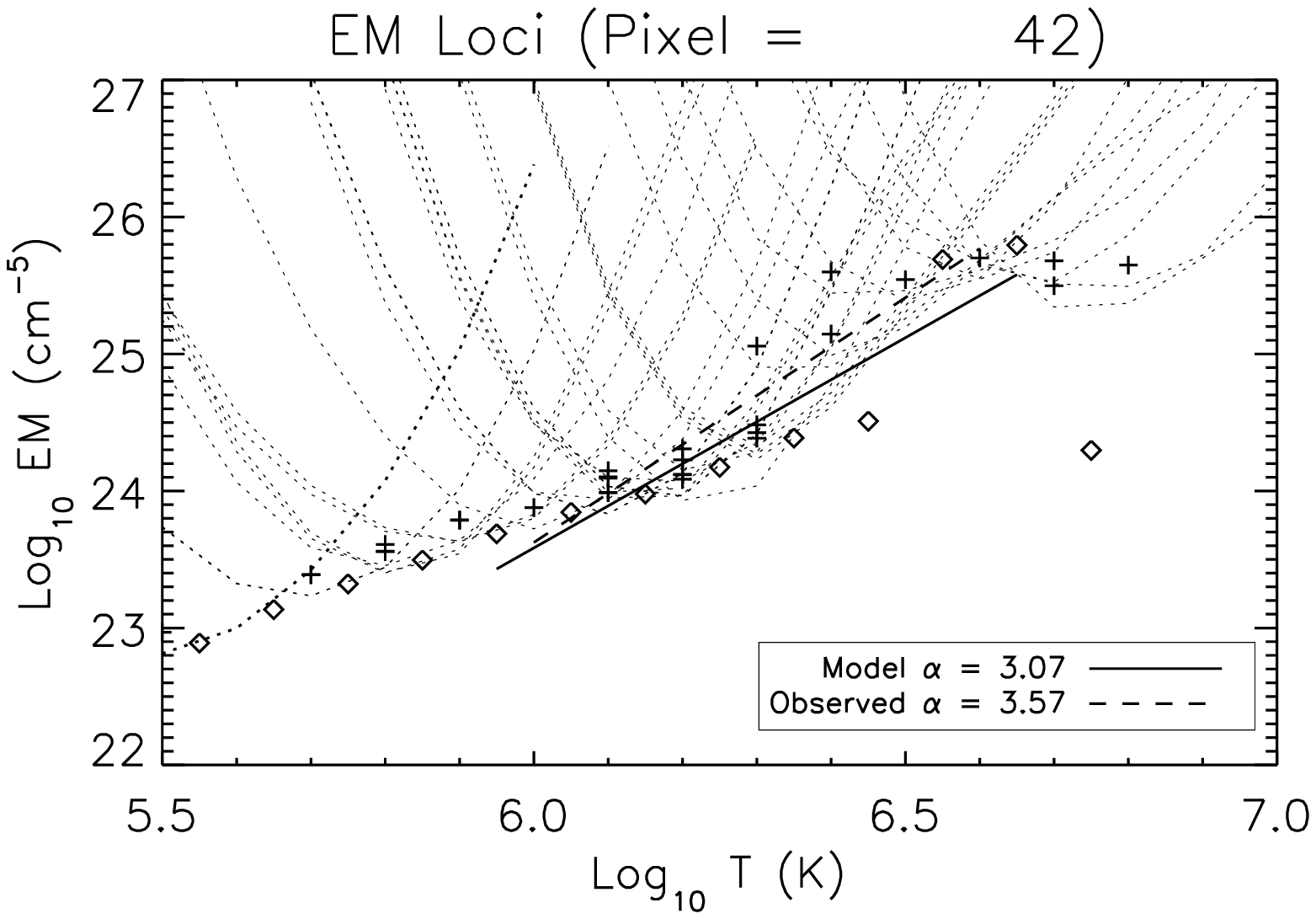}
\caption{Results for Run~40.}
\label{fig6}
\end{figure}

Now, examine Figure~\ref{fig6} for Run~40 ($N=20$, $\tau_H=300$~s, $\tau_C=300$~s), which is at the opposite extreme to Run~21 for the set of Runs where $2L=80$~Mm. It can be seen from the upper panels that the coronal density now reaches and maintains its peak during the nanoflare train, which means that we can expect a significant contribution to the EM from the heated strands. The EM shown in the lower panel confirms our expectation. The temperature of the hot strands varies between 3.5~MK and 5~MK, with an average of 4.25~MK, and the peak of the EM is substantially enhanced between between  $10^{6.6}$~K and $10^{6.7}$~K and $\log_{10} \left( 4.25 \times 10^6 \right) = 6.62$. The temperature variation allowed by $\tau_C$ during the nanoflare train is $\Delta \log_{10} T = \log_{10} \left( 5 \times 10^6 \right) - \log_{10} \left( 3.5 \times 10^6 \right)=0.15$ and, since the EM is binned in temperature intervals of 0.1~dex, this is sufficient for the neighboring bins of the peak to also be populated. In consequence, the hot component of the EM is not quite isothermal and it may be possible to diagnose the parameter $\tau_C$ from its width (as we show later in this Section). The EM in the region of the peak is therefore strongly enhanced relative to the part of the EM that is due primarily to the cooling (post-nanoflare train) strands. Compare this with Figure~\ref{fig3} in which no significant enhancement is apparent. It is straightforward to see that a linear regression applied between the temperature of peak EM and 1~MK in Figure~\ref{fig6} will yield a steeper slope than the same analysis carried out for the EM shown in Figure~\ref{fig3}.

As long as $\tau_C < \Delta_C$, the EM of the cool component is set by $\Delta_C$ and the EM of the hot component is set by $\Delta_H$, which in turn depends on $N$, $\tau_H$ and $\tau_C$ ($\Delta_H=N\tau_H+(N-1)\tau_C$). A longer timescale $\Delta_H$ means that a greater proportion of the total number of loop strands are at some stage of the nanoflare train. This enhances the part of the EM that is determined by the nanoflare train relative to the part that is cooling following the cessation of the train. From the perspective of an individual strand, the plasma spends more time in the hot state and this translates into a larger EM. We can steepen the EM slope by changing $N$, $\tau_H$ or $\tau_C$ to increase $\Delta_H$, but $\Delta_C$ is essentially out of our control because it depends primarly on the length of the loop. If $\tau_C > \Delta_C$, the situation reverts to that of low-frequency nanoflares where the slopes are shallower.

\begin{figure}
\centering
\includegraphics[width=0.8\textwidth]{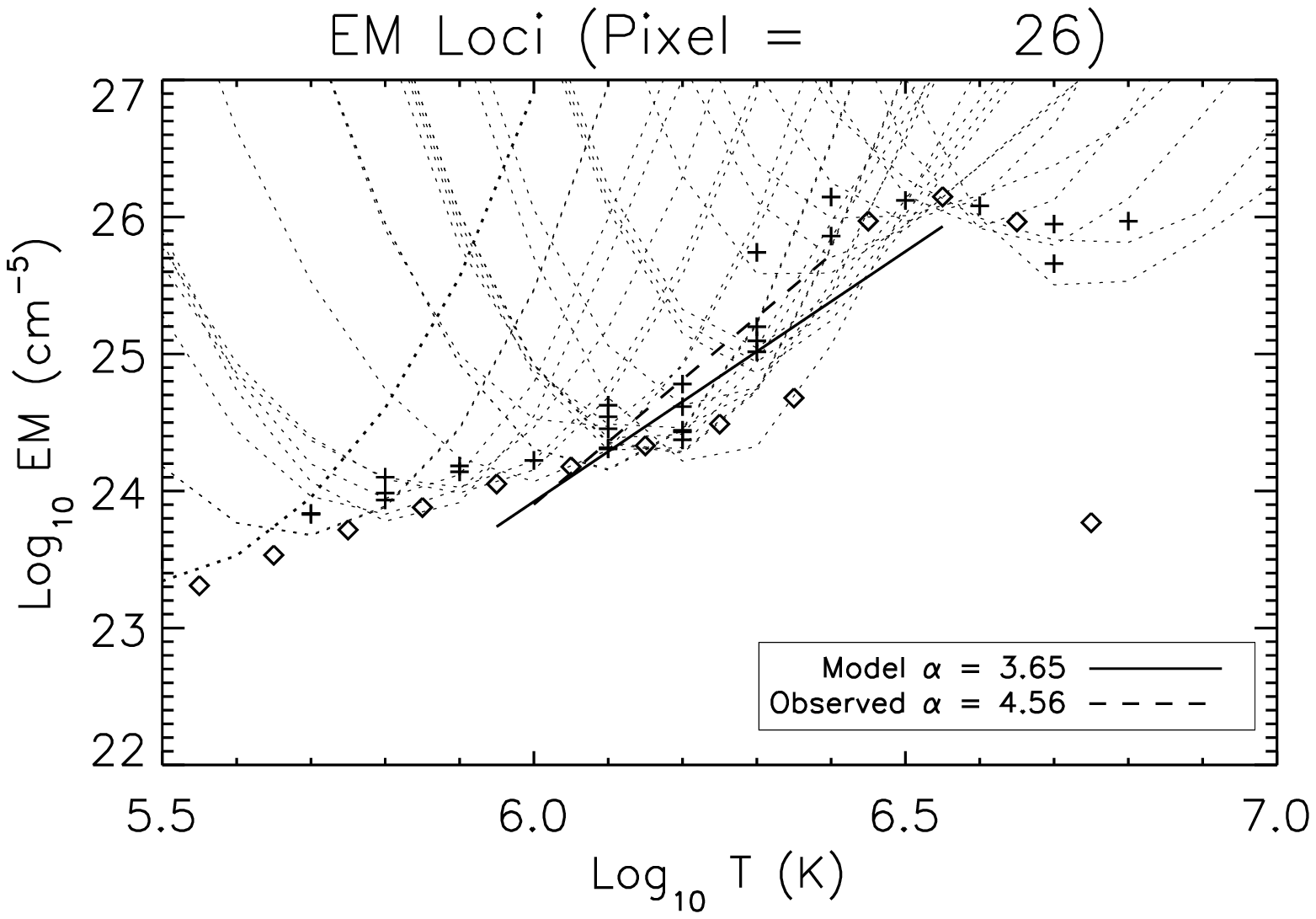}
\includegraphics[width=0.8\textwidth]{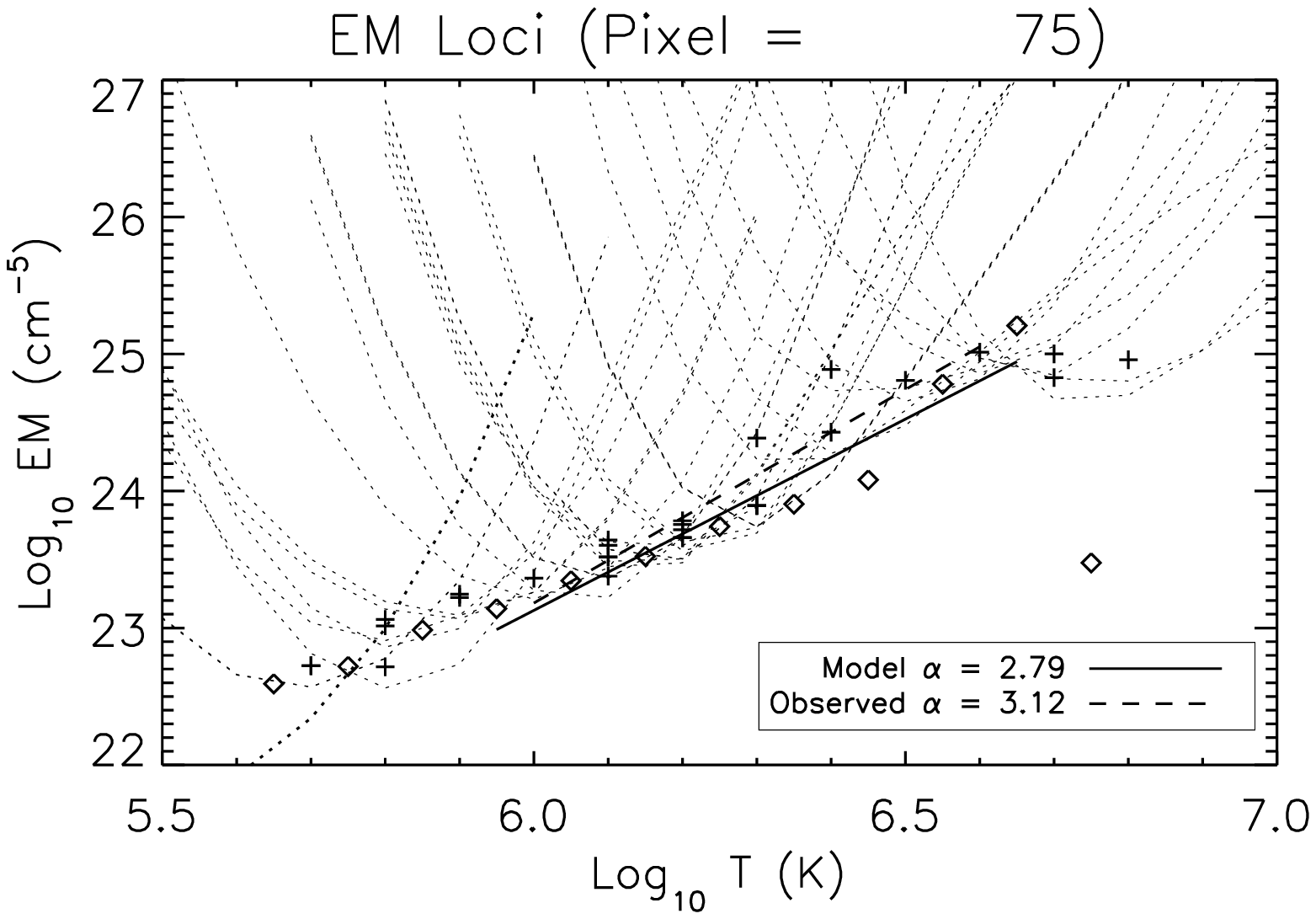}
\caption{The emission measure distributions for Runs~20 (upper-panel) and 60 (lower-panel).}
\label{fig7}
\end{figure}

We can now understand why longer loops tend to yield shallower EM slopes between the temperature of peak EM and 1~MK. The cooling time $\Delta_C$ tends to be greater for longer loops and for the fixed range of $\Delta_H$ that we consider, the ratio $\Delta_H/\Delta_C$ is generally smaller. Following the interpretation described above; a smaller proportion of the total number of strands will now be at some stage of the nanoflare train and so the hot part of the EM will not be so enhanced relative to the post-train part of the EM.  Figure~\ref{fig7} shows the EMs for Runs~20 ($2L=40$~Mm) and 60 ($2L=160$~Mm), where $N=20$, $\tau_H=300$~s, $\tau_C=300$~s in both cases. It can be seen that the hot part of the EM is greatly enhanced relative to the cool part in Run~20, whereas the enhancement isn't so pronounced in Run~60. In the case of Run~60, the most extreme of the $2L=160$~Mm set of Runs, the ratio $\Delta_H/\Delta_C=1.80$ compared to values over 3 for shorter loops. The enhanced hot part of the EM clearly leads to steeper slopes and, therefore, in order to obtain steeper slopes for longer loops the quantity $\Delta_H$ must be increased relative to $\Delta_C$.

\begin{table}
\centering
\caption{Distribution of observed emission measure slopes cool-ward of the peak.}
\begin{tabular}{c c c c}
\tableline
 & $\alpha - \Delta \alpha$ & $\alpha$ & $\alpha + \Delta \alpha$ \\
\tableline
$\alpha \le 1.0$ & 3 & 0 & 0 \\
$1.0 < \alpha \le 1.5$ & 5 & 0 & 0 \\
$1.5 < \alpha \le 2.0$ & 3 & 3 & 0 \\
$2.0 < \alpha \le 2.5$ & 6 & 5 & 0 \\
$2.5 < \alpha \le 3.0$ & 2 & 3 & 3 \\
$3.0 < \alpha \le 3.5$ & 2 & 6 & 5 \\
$3.5 < \alpha \le 4.0$ & 1 & 2 & 3 \\
$4.0 < \alpha \le 4.5$ & 0 & 2 & 6 \\
$4.5 < \alpha \le 5.0$ & 0 & 1 & 2 \\
$\alpha > 5.0$ & 0 & 0 & 3 \\
 & 100\% & 100\%  & 86\% \\
\end{tabular}
\label{table4}
\end{table}

Table~\ref{table4} shows the distribution of observed EM slopes cool-ward of the peak, $1.70 \le \alpha \le 5.17$ (Table~3 of Paper~I), where the uncertainty in the slope $\Delta \alpha=\pm1.0$ (Guennou et al. 2012a,b; Paper~I). Given our predicted range of $0.88 \le \alpha_{\mbox{observed}} \le 4.56$ and the range of uncertainty, we can see that the nanoflare train heating scenario is consistent with 86\% to 100\% of the observed range of EM slopes. This may be compared with 0\% to 77\% in the low-frequency nanoflare scenario found in Paper~I. We encourage the gathering of many more observational measurements of EM slopes from large-scale surveys of active regions in order to improve these statistical estimates. Furthermore, EM reconstructions applied to a far larger set of real observational data, such as would be obtained from large-scale surveys, would directly address the question of the ubiquity of a hot component to the EM and thereby provide direct evidence either for or against the heating scenario described here.

Given a sufficiently large sample of EM observations it may be possible to estimate $\tau_C$ from the characteristic width of the hot component in the following way. The cooling timescales by thermal conduction and radiation are given by

\begin{equation}
\tau_{\mbox{cond}} = 4\times10^{-10} \frac{nL^2}{T^{5/2}}~\mbox{~~~~and~~~~}~\tau_{\mbox{rad}} = \frac{3k_B}{\chi n T^{b-1}},
\label{eqn3}
\end{equation}

\noindent where $\chi$ and $b$ are given by piece-wise power-law fits to the curve of the radiative loss rate as a function of temperature \citep[e.g.][]{klimchuk2008}, and the ratio of these cooling timescales is given by

\begin{equation}
\frac{\tau_{\mbox{cond}}}{\tau_{\mbox{rad}}} = \left(9.66\times10^5\right) \chi \left(nL\right)^2 T^{b-7/2},
\label{eqn4}
\end{equation}

\noindent which can be calculated from an observed EM where $T=T_{\mbox{peak}}$ if $L$ and $n$ are known. The temperature and density range during the nanoflare train in Figure~\ref{fig6}, together with the loop length $2L=40$~Mm and the values of $\chi$ and $b$ given by \cite{klimchuk2008}, yields $\tau_{\mbox{cond}}/\tau_{\mbox{rad}}=0.03$. We conclude that thermal conduction is the dominant cooling mechanism during the interval $\tau_C$, in this example, and therefore primarily responsible for setting the width of the hot component. Using the temperature evolution in the case of non-evaporative conductive cooling (since the density does not change much during relatively long trains) given by \cite{antiochos1976}

\begin{equation}
T(t) = T_0 \left(1 + \frac{t}{\tau_{\mbox{cond}}} \right)^\beta,
\label{eqn5}
\end{equation}

\noindent where $\beta = -2/5$, we find

\begin{equation}
\frac{dT}{dt} = \frac{\beta T}{\tau_{\mbox{cond}} + t}.
\label{eqn6}
\end{equation}

\noindent We can approximate the width of the hot component $\Delta T$ by the change in temperature during the interval $\tau_C$.

\begin{equation}
\frac{\Delta T}{\tau_C} \approx \frac{\beta T}{\tau_{\mbox{cond}} + \tau_C}.
\label{eqn7}
\end{equation}

\noindent Since the temperature is usually binned in units of $\log_{10} T$  we can write

\begin{equation}
\Delta \log_{10} T \approx \frac{\beta \tau_C}{\left(\ln 10\right) \left(\tau_{\mbox{cond}} + \tau_C\right)}.
\label{eqn8}
\end{equation}

\noindent The quantity $\Delta \log_{10} T < 0$ because the plasma is cooling. Finally, the interval $\tau_C$ can be found in terms of the approximate width of the hot component from

\begin{equation}
\tau_C \approx \tau_{\mbox{cond}} \left( \frac{\beta}{\left(\ln 10\right) \left(\Delta \log_{10} T\right)} - 1 \right)^{-1}.
\label{eqn9}
\end{equation}

\noindent When $\tau_C >> \tau_{\mbox{cond}}$ we note from Equation~\ref{eqn8} that $\Delta \log_{10} T = -2/5(\ln 10)$, which sets an upper limit to the component of the width of the hot part of the EM that can be due to cooling by thermal conduction. To diagnose $\tau_C$ in this regime using Equation~\ref{eqn9}, temperature bins that are substantially narrower than this limit of $|\Delta \log_{10} T| \approx 0.17$~dex are required and we conclude that the typical bin width of 0.1~dex is too coarse to be useful.

In the regime where $\tau_{\mbox{cond}}/\tau_{\mbox{rad}}>1$ radiative cooling dominates during the interval $\tau_C$. Using the temperature evolution in the case of radiative cooling given by \cite{cargill1995}

\begin{equation}
T(t) = T_0 \left(1 - (1-b)\frac{t}{\tau_{\mbox{rad}}} \right)^\frac{1}{1-b}
\label{eqn10}
\end{equation}

\noindent and following a similar analysis to that presented for cooling by thermal conduction, we find

\begin{equation}
\Delta \log_{10} T \approx \frac{\tau_C}{\left(\ln 10\right)\left(\left(1-b\right)\tau_C - \tau_{\mbox{rad}}\right)}
\label{eqn11}
\end{equation}

\noindent and

\begin{equation}
\tau_C \approx \tau_{\mbox{rad}} \left( (1-b) - \frac{1}{\left(\ln 10\right) \left(\Delta \log_{10} T\right)} \right)^{-1}.
\label{eqn12}
\end{equation}

\noindent When $(1-b)\tau_C >> \tau_{\mbox{rad}}$ we note from Equation~\ref{eqn11} that $\Delta \log_{10} T$ is a positive quantity and yet we require it to be negative for cooling. The only way to satisfy this constraint is to ensure that $(1-b)\tau_C < \tau_{\mbox{rad}}$ and this implies $\tau_C < \tau_{\mbox{rad}} / (1-b)$, which sets the upper limit to the inter-nanoflare timescale $\tau_C$ that can be diagnosed using Equation~\ref{eqn12}. For example, $b=-1/2$ is commonly used in analytical approximations to the radiative loss curve in the range $\log_{10} T > 5.0$ \citep{bradshaw2010} and this sets the limit $\tau_C \rightarrow 2\tau_{\mbox{rad}}/3$ as $\Delta \log_{10} T \rightarrow \infty$. In this regime, we conclude that temperature bins of 0.1~dex may be adequate depending upon the relative values of $\tau_C$ and $\tau_{\mbox{rad}}$. Following a different approach, it may also be possible to constrain the parameter space of $N$, $\tau_H$ and $\tau_C$ from time-series studies of active region emission across a range of temperatures, such as those carried out by \cite{viall2011,viall2012}.

Figures~\ref{fig4} and \ref{fig5} show that $\alpha_{\mbox{observed}}$ and $\alpha_{\mbox{model}}$ diverge as the EM slope steepens, such that the slope of the EM derived from synthetic Hinode-EIS data increasingly over-estimates the slope of the model EM (although in many cases each slope value falls within the error bars of the other one). This is particularly noticeable for $\alpha>3$. It may be attributed to the estimate of the electron density used in the contribution function $G(n,T)$ to derive the loci-curves ($EM(T) \propto I / G(n,T)$). The line intensities in our forward model are computed using contribution functions that are density dependent. However, when calculating the EM using the Pottasch method the line intensities are divided by contribution functions that are no longer density dependent (assuming a fixed value of $n=10^9$~cm$^{-3}$). The hotter strands have the highest densities, but fixing the contribution function density at $10^9$~cm$^{-3}$ for density sensitive lines leads to an underestimate of the true contribution function and an overestimate of the EM at higher temperatures. The density sensitive lines therefore introduce a bias toward steeper slopes. To mitigate this effect we recommend that density measurements accompany EM calculations, which provides another good reason for measuring $n$ in addition to needing it for estimates of $\Delta_C$.

The shape of $G(n,T)$ can also play a role in the discrepancy between the model and predicted slopes because the Pottasch method assumes that $G(n,T)$ is constant in the range $\Delta \log_{10} T = \pm0.15$ to either side of the peak formation temperature of the emission line, and zero everywhere else. The constant value is equal to the average of the actual $G(n,T)$ in this range. If two emission lines with the same peak formation temperature have $G(n,T)$'s with appreciably different shape (e.g. one is substantially narrower than the other) then this could also affect the more density sensitive lines, leading to a discrepancy between the model and predicted values.

\section{Summary and Conclusions}
\label{summary}

We have run a series of numerical experiments to explore coronal loop heating by nanoflare trains with the intention of extending our previous study, presented in Paper~I, of heating by low-frequency nanoflares. The aim of the present study is to determine whether nanoflare trains are able to overcome the intrinsic limit to the EM slopes (from the EM peak to 1~MK) that we demonstrated in our previous work \citep[see also:][]{mulumoore2011} so that observationally measured slopes of $\alpha > 2.6$ can be explained. By constructing EMs from synthetic Hinode-EIS data, using spectral lines formed over a wide range of temperatures, we predict slopes in the range $0.88 \le \alpha_{\mbox{observed}} \le 4.56$ for the nanoflare train heating scenario encapsulated by the extent of the parameter space that we have chosen for our study.

Though we have not attempted to match any specific observations in the work presented here, we have reached a set of conclusions concerning the broad properties that the nanoflare trains must possess to be consistent with observations. A number of our Runs yield temperatures of the peak EM that are probably too low to agree with the observed range in active region cores; for example, Runs~[5-8,21,41,45] have $\log_{10} T_{\mbox{peak}} = 6.35$. We may therefore exclude the particular combination of parameter values that led to this result for our fixed value of $E_{H0}$, which are in general: small $N$; short $\tau_H$; and long $\tau_C$. Stronger heating would boost $\log_{10} T_{\mbox{peak}}$ into the required range, but must of course remain consistent with the free magnetic energy.

In the case of loops of fixed length $2L$ we find steeper EM slopes with increasing $N$, $\tau_H$ and $\tau_C$. For fixed $E_{H0}$ the parameters $N$ and $\tau_H$ determine the total energy released during the nanoflare train and $\tau_C$ determines the width of the hot component of the EM. We have parameterized the relationship between [$N,\tau_H,\tau_C$ ] and $\alpha_{\mbox{observed}}$ using the single parameter $\Delta_H/\Delta_C$, which is the ratio of the number of strands in some stage of nanoflare train heating to the number of strands undergoing cooling and draining following the cessation of the nanoflare train (Figures~\ref{fig4} and \ref{fig5}). We found that the general trend is for $\alpha_{\mbox{observed}}$ to increase with $\Delta_H/\Delta_C$. The physical reason for this is that the EM of the cool component is set by $\Delta_C$ and the EM of the hot component is set by $\Delta_H=N\tau_H+(N-1)\tau_C$. A longer timescale $\Delta_H$ means that a greater proportion of the total number of loop strands are in some stage of heating. This enhances the part of the EM that is determined by the heating relative to the part that is cooling and draining.

In the case where $2L$ is allowed to vary we find it relatively easy to obtain slopes consistent with the upper-range of those observed for shorter loops (e.g. $2L=40$~Mm, Figure~\ref{fig4}), within the parameter space of nanoflare train properties that we explored. The range of slopes for longer loops ($2L \ge 80$~Mm) that we found within this parameter space are largely consistent with low-frequency nanoflares. All but two values of $\alpha_{\mbox{observed}}$ are consistent with low-frequency nanoflares in the case of $2L=160$~Mm. However, we also know from our investigations that obtaining steeper slopes for longer loops is a matter of allowing the ratio $\Delta_H/\Delta_C$, and therefore the quantity $\Delta_H$, to increase. Given an observationally measured EM slope of $\approx 3$, Figures~\ref{fig4} and \ref{fig5} show that the expected ratio $\Delta_H/\Delta_C \approx 1.5-2.0$ but we cannot determine $\Delta_H$ uniquely from this information alone. However, given the length and density (from spectral line ratios) of the structure then it is possible to estimate $\Delta_C$ as the radiative and enthalpy-driven cooling timescale \citep[e.g.][]{bradshaw2010} from the temperature of peak emission measure (a known quantity from the EM plots). Alternatively, it may be possible to measure $\Delta_C$ by following the cooling structure via emission from recombining ions or successively cooler wavelength channels. We urge observers to include such estimates ($2L$, $n$ and~/~or $\Delta_C$) along with all measurements of EM slopes, in order that possible heating scenarios may be more rigorously tested against observational data.

We have found that the properties of 86\% to 100\% of active region cores can be consistent with heating in the form of nanoflare trains, where we have assumed that not all of the free magnetic energy is released at once, but rather in a series of individual nanoflares that we refer to as a train. However, there are alternative scenarios. In one example, all of the energy available for heating may be released and the magnetic stress then builds up again in the interval between nanoflares. In a modeling sense this is similar to the scenario that we have explored here, but with a smaller amount of free energy in the field before the onset of heating because it only has to supply a single event before recharging. The parameter $\tau_C$ then becomes a recharging timescale for the magnetic field. Suppose that the storage of energy in the field is a consequence of the field lines becoming twisted and tangled due to convective motions on the surface and the energy is released when some critical condition is satisfied \citep[e.g. the angle between neighboring field lines exceeds some critical value:][]{parker1988,dahlburg2005}. If the surface motions do not vary too strongly then it is reasonable to suppose that the energy stored in the magnetic field, and ultimately released, be proportional to the recharging timescale. We cannot predict when reconnection will occur, but we can place constraints on this timescale. It cannot be so short that the heating is effectively steady, because this would give rise to a more or less isothermal emission measure. Conversely, for EM slopes greater than about 2.6 it cannot exceed a cooling timescale because this would be the limit of low-frequency nanoflare heating in which slopes are constrained to a range of values less than 2.6. The recharging timescale must be of just the right length such that the width of the hot component of the EM is consistent with what is observed.

Previous studies of the relationship between the EM and the timescales related to the heating mechanism have led to findings that are consistent with ours, though for conceptually somewhat different scenarios. For example, \cite{testa2005} explored a parameter space of heating pulses with timescales that are longer ($300-1100$~s) than those we have chosen, though the total duration of heating is similar, and with the energy release localized at the loop foot-points. When the spatial scale of heating is sufficiently small a thermal instability arises, leading to an ongoing cycle of heating and cooling which gives rise to the warm component of the EM in their study. Their cooler loops reach similar temperatures ($T = 10^{6.48}$~K) to some of ours, but their hotter loops reach significantly higher temperatures ($T=10^7$~K) due to stronger heating, and they adopted a loop length of $2L=200$~Mm that is some 40~Mm longer than our upper limit. They find EM slopes approaching 5.0, which are steeper than the slopes we find for our $2L=160$~Mm loops but, as we have already discussed, extending $\Delta_H$ relative to $\Delta_C$ (consistent with estimates of the free magnetic energy) will result in steeper EM slopes. In the case of \cite{testa2005} their EM slopes are determined by the relative lengths of the heating and cooling phases of the cycle that are ultimately governed by the timescales associated with the evolution of the thermal instability.

It is abundantly clear that different sets of observational data to complement EM slope and related measurements are required to help constrain the timescales on which the active region heating mechanism operates, and ideally to help constrain the mechanism itself. One such data set may include the findings of \cite{ugarteurra2012}, which indicate that active region variability evolves over time as do the relative contributions to the EM of hot and warm plasma. This suggests that the frequency of heating may also change, with older ARs heated by higher-frequency events, and provides a likely explanation for why \cite{warren2011} found that high-frequency heating was most consistent with their observations, whereas others have found that low-frequency heating is most appropriate in other cases (e.g. Paper~I). An alternative data set may be found in the temperature dependent pattern of red- and blue-shifts that are observed in active regions. An extension to the work presented here will be an analysis of the Doppler-shifts as a function of temperature associated with low-frequency and nanoflare train heating in order to determine whether they are consistent with the patterns of flows that are observed in active regions, and to investigate their potential to provide additional constraints on the parameter space of heating properties. This will be the subject of the next paper in this series.

\acknowledgments

SJB and JAK acknowledge support for this work by the NASA SR\&T program. We thank the International Space Science Institute (ISSI) for hosting the International Team led by SJB and Helen Mason, and the team members for the fruitful discussions that took place during the meeting held there in February 2012. Our thanks to the referee for their comments and suggestions which improved the original manuscript.

\appendix

\section{Spectral Line Synthesis}
\label{appA}

The calculation of the total (wavelength integrated) intensity of a line in instrument units (DN~pixel$^{-1}$~s$^{-1}$) is described in detail by \cite{bradshaw2011} and summarized for convenience here:

\begin{equation}
I = {\mbox{IR}}(\lambda) \times G(\lambda,n,T) \times <EM>,
\label{eqnA1} 
\end{equation}

\noindent where ${\mbox{IR}}(\lambda)$ is the instrument response function (units: DN~pixel$^{-1}$~photon$^{-1}$~sr~cm$^2$), which is the product of the effective area, plate scale, and gain of the instrument.  The Hinode-EIS response functions are available in the SolarSoft package.

$G(\lambda,n,T)$ is the contribution function, defined as:

\begin{equation}
G(\lambda,n,T) = \frac {0.83 \times Ab(Y) \times Y_i \times \epsilon(\lambda,n,T)} {4\pi \times (hc/\lambda)},
\label{eqnA2}
\end{equation}

where 0.83 is the proton:electron ratio, $Ab(Y)$ is the abundance (relative to hydrogen) of element $Y$, $Y_i$ is the population fraction of charge state $i$ of element $Y$, $\epsilon(\lambda,n,T)$ is the emissivity of the line (units: erg~s$^{-1}$~cm$^3$) calculated using the EMISS\_CALC function \citep[Chianti:][]{dere1997,dere2009}, and $hc/\lambda$ is the photon energy (units: erg).

$<EM>$ is the spatially averaged column emission measure in the pixel (units: cm$^{-5}$):

\begin{equation}
<EM> = \frac{EM_{\mbox{pix}}}{A_{\mbox{pix}}},
\label{eqnA3}
\end{equation}

\noindent where $EM_{\mbox{pix}} = n_e^2 d\!V$ is the emission measure in the pixel and $A_{\mbox{pix}}$ is the pixel area. It is straightforward to see how the emission measure can be calculated from the line intensity, given the contribution function and the instrument response.

The intensity per unit wavelength of the spectral line is given by convolving the total intensity with the line broadening function:

\begin{equation}
I_\lambda = \frac{I}{\sqrt{\pi \sigma^2}} \exp \left( - \frac{\left( \lambda - \lambda_0 - \Delta \lambda \right)^2}{\sigma^2} \right),
\label{eqnA4}
\end{equation}

\noindent where $\lambda_0$ is the rest wavelength of the line and $\Delta \lambda$ is the Doppler shift due to the line-of-sight component of the bulk flow, given by $\Delta \lambda = \frac{\lambda_0}{c} v$. The total line width, $\sigma$, is determined by the thermal width and the instrument width:

\begin{equation}
\sigma^2 = \frac{2}{3} \sigma_{\mbox{th}}^2 + 0.36 \sigma_{\mbox{ins}}^2,
\label{eqnA5}
\end{equation}

\noindent where $\sigma_{\mbox{th}} = \frac{\lambda_0}{c} v_{\mbox{th}}$ and suitable values for $\sigma_{\mbox{ins}}$ are 60~m\AA~(EIS short wavlength channel) and 67~m\AA~(EIS long wavelength channel). In calculating the spectral line profile one must be careful to ensure that $I = \int I_\lambda d\lambda$.

The line intensity from each cell of the numerical grid are then added to the total for the appropriate detector pixel(s). The correct pixel is determined by projecting each grid cell onto a single row of detector pixels and establishing what proportion of the emission from the grid cell falls onto each pixel \citep[see Figure~1 of][]{bradshaw2011}. This procedure is repeated for all of the spectral lines of interest, for all of the grid cells of the strand, and for every strand comprising the loop. The result is then a prediction of the total emission along the loop, due to the contributions from the many sub-resolution strands at different stages of evolution, as would be measured by a particular observing instrument.

\end{document}